\documentclass[a4paper,titlepage,11pt,a4]{article}

\usepackage[dvips]{graphicx}
\usepackage{listings}
\usepackage{ulem}
\usepackage{subfigure}
\usepackage{amsfonts}
\usepackage{url}

\include{epsf}

\begin{document}

\begin{titlepage}
\begin{center}
\LARGE Unstable Optical Resonators \& Fractal Light\\
\vskip 1cm
\includegraphics[width=10cm]{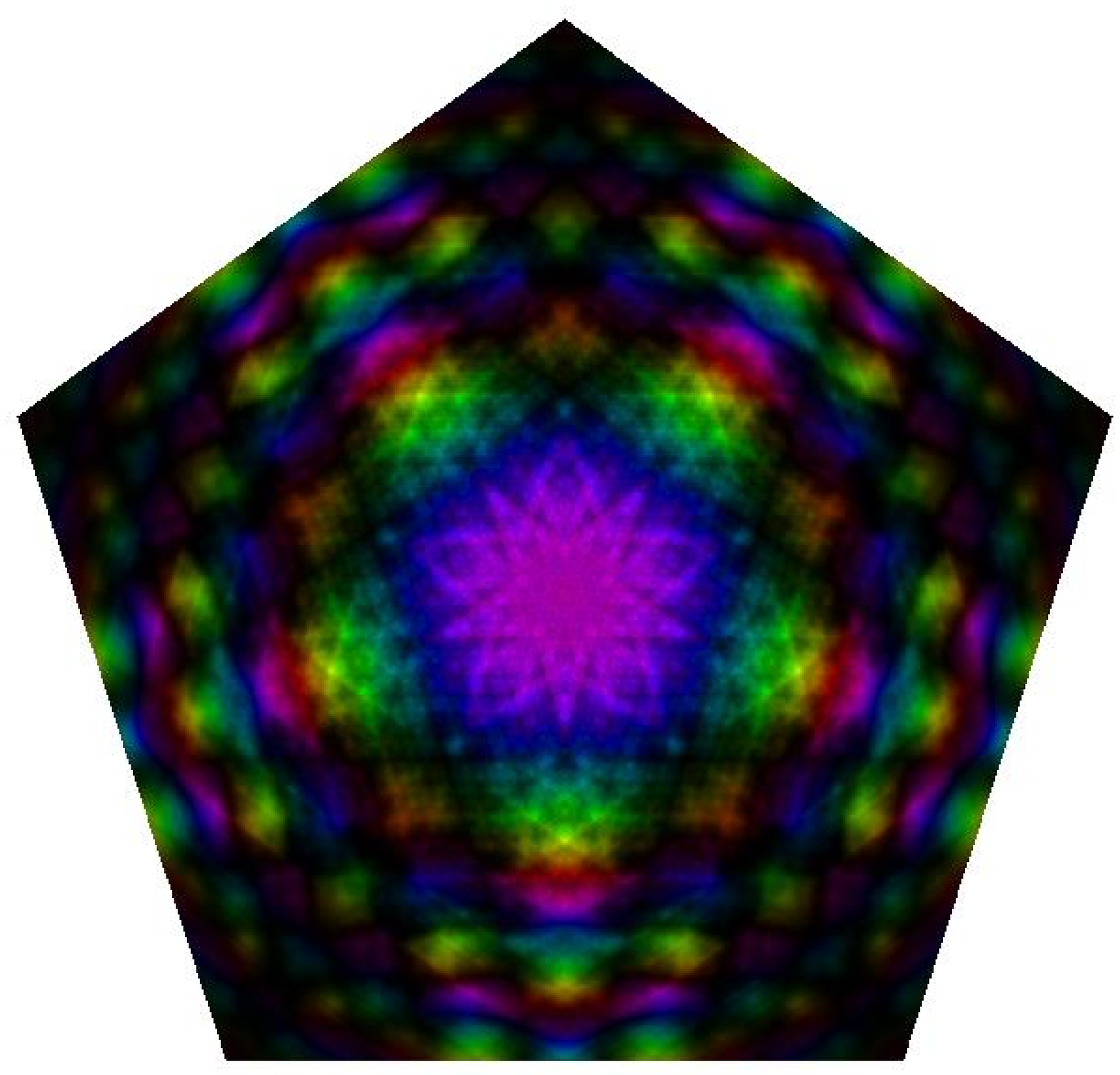}
\\
\large MSci Project Report
\vskip 5mm
\large Jarvist Moore Frost
\\ (email: {\tt jarvist.frost@ic.ac.uk})
\vskip 5mm
\today
\end{center}
\end{titlepage}

\pagenumbering{roman} \setcounter{page}{1}
\section*{Abstract}
Codes were written to simulate the propagation of monochromatic light
through a bare optical resonator, using a computational Fourier method to
solve the Huygens-Fresnel integral. This was used, in the Fox-Li method, to
find the lowest-loss eigenmodes of arbitrary cavity designs. An
implicit shift `hopping' method was employed to allow a series of
increasingly higher-loss eigenmodes to be found, limited in number by
computational time.

Codes were confirmed in their accuracy against the literature, and were used
to investigate a number of different cavity configurations.

In addition to confirming the fractal nature of eigenmodes imaged at the
conjugate plane of a symmetric ($g<-1$) resonator, 
an initial study was made of how the (imperfect) quality of the
fractal fit varied as the defining aperture was moved around the cavity.

A comparison was also made with the fractal-patterns produced
by codes written to simulate basic video-feedback.
\newpage

\section*{Acknowledgements}
Many thanks to my supervisor, Professor Geoff New, for the sheer amount
of time he invested in an undergraduate project, specially prepared notes 
\& loans of reference books.
Perhaps more importantly - he
was always upbeat no matter how slow progress was going, and took
a very philosophical attitude towards bugs \& setbacks!

Thanks to Benjamin Hall, my lab partner, for sharing the work-load, 
helping me get up to speed with geometric optics at the start of
the year and discussions over an uncountable number of cups of tea and
reams of scribbled diagrams.
\newpage

\tableofcontents
\newpage

\listoffigures
\newpage

\pagenumbering{arabic} \setcounter{page}{1}
\section{Theoretical Groundwork}
\subsection{Optical Resonators}
The simplest optical resonator consists of two curved mirrors set up facing
each other. We can define some useful quantities known as g-factors that 
define the behaviour of the 
beam in terms of spacing and radius of curvature of
the mirrors [Eqn. \ref{gfactor}]. These are a scale independent way of
defining the behaviour of the cavity.

\begin{equation} \label{gfactor}
g_1 \equiv 1 - \frac{L}{R_1} \qquad g_2 \equiv 1 - \frac{L}{R_2}
\end{equation}

\subsection{Stability Condition}

A stable cavity is one in which (by a simple geometric argument) any light ray 
in the cavity will be trapped between the mirrors and redirected towards the 
centre of the cavity.

This condition can be easily defined in terms of the g-factors 
[Eqn. \ref{gstability}] and gives
rise to a region on a $g_1 vs. g_2$ plot that corresponds to stable
cavity configurations. 

\begin{equation} \label{gstability}
0 \le g_1 g_2 \le 1
\end{equation}

When the g-factors exceed this region, a round-trip of the cavity becomes
overall magnifying, and
light inevitably `spills' over the edge of the mirrors.

\subsection{Transverse Eigenmodes}

By considering a short pulse or `slab' of radiation indefinitely
recirculating between the mirrors of the cavity, and demanding that the
transverse beam pattern be unchanged by propagation (i.e. a stable `lasing'
beam) one can identify a transverse pattern of the field that supports 
such indefinite oscillations without changing. This is known as the
Transverse Eigenmode.

The action of the optical resonator can be considered as a matrix operator
$\uuline{M}$ which modifies the transverse profile of the wavefront $U$.
This leads to a basic eigenvalue equation:

\begin{center}
$\uuline{M} U_n = \gamma_n U_n$
\end{center}

Finding the eigenmodes of an optical resonator is simply a matter of solving
this equation. However, $\uuline{M}$ is not a simple linear operator but
corresponds to the physical evolution of light (including diffraction
effects) as it passes around the cavity.

These $U_n$ show a certain transverse (across the cavity)
distribution that is very strongly dependent on varying cavity design. 
There are many such eigenmodes for a given cavity. The higher the
order mode, the greater the round-trip loss of energy.

\subsection{Geometrically Stable Cavities \& Hermite-Gaussian Modes}

In stable optical resonators, the eigenmodes produced are
essential plane waves, multiplied by a transverse mode function. When
expanded in rectangular transverse coordinates, these modes are given almost
exactly by Hermite-Gaussian functions\cite{siegman}.These are often referred 
to as $TEM_{n,m}$ waves\cite{siegman}.

The transverse distribution of light intensity resembles a combination of Gaussian
distributions.  Therefore the majority of
energy in the beam is concentrated in only a small region of the
gain medium, leaving most of the gain underutilised.

\subsection{Planar Resonators}

The planar resonator occupies the [1,1] point of the g-factor stability diagram,
and as such it might be considered somewhat surprising that stable
eigenmodes exist on this boundary. 
The analytic treatment is far more difficult than for stable cavities, but in 
appearance the Eigenmodes are very similar to the $TEM$ modes, with the addition
of Bessel functions adding diffraction ripples. 

Unlike in a confocal resonator, these modes do not go to zero at the mirror 
edges, and a quantity of light is lost around the edges.

\subsection{Unstable Cavities}

Due to the positive magnification on each round trip of the cavity, a large
proportion of the energy is lost past the mirror - in
fact this can form a useful `doughnut' output beam. Analytic expressions for
the mode patterns are difficult to find, and it is
these cavity designs which we used for the majority of our project.

Unstable Cavities have a number of benefits for high-power laser systems,
including far better utilisation of the bulk of the gain-medium.

\subsection{Virtual Source Technique}

By unfolding the mirrored cavity into a series of effective apertures and
lenses, the modal pattern can be inferred from taking a weighted sum of the
edge diffraction effects \& the non-diffracted plane wave passing
through\cite{pLasers}.

However, this technique has been mainly developed for square or $1D$ slit
apertures\cite{kaleidoscope}, and so with the far more complicated mirror shapes that we hope to
study, we will be forced to take an entirely numerical approach.

\subsection{Huygens' Integral}

By considering the propagation of a light beam as the combination \&
interference of spherical Huygens' wavelets originating at the source, 
the light field at any observable location can be derive. 
By considering a paraxial beam, 
where the distance between
input \& output planes is considered large enough so that  $cos \theta
\approx 1$ and utilising a simplified Paraxial-Spherical form of
propagating wavelets,
one can form the Fresnel approximation to Huygens' integral\cite{siegman}.

This is made further useful for us by separating it into two one-dimension
integrals [Eqn. \ref{hfint}]\cite{siegman}.

\begin{equation} \label{hfint}
\tilde{u}(x,y)= \sqrt \frac{j}{L\lambda} \int {\tilde{u}_0(x_0\,z_0) exp (-j
\frac{\pi(x-x_0)^2}{L\lambda}) dx_0}
\end{equation}

This form of the integral represents a convolution of the input field
$\tilde{u}_0$ with a spherical wave-function 
$exp(-j\pi x_{0}^{2} / (z-z_0)\lambda)$. 
Convolution can be easily achieved by Fourier-transforming
the two functions, doing a product multiplication of the results, then
inverse Fourier transforming to produce the desired convolution\cite{siegman}.

This process can then be repeated for a given laser cavity with suitable
interacting gain medium, until convergence on a `steady state' modal pattern
is observed\cite{1975} which can be considered an eigenmode of the given
cavity.

\subsection{Ray Matrices}
Within the paraxial treatment, the effect of a number of optical elements
(such as thin lenses, mirrors etc.) can be considered as items that modify
the slope \& displacement of incoming light beams by linear transformation.
Therefore, one can define a rank 2 tensor (ray matrix) that acts upon a ray
vector to transform it into what is present after the optical element.

When multiple elements are cascaded together, an overall ray matrix can be
calculated by simply taking the normal matrix product of the constituent
elements, algebraically arranging the ray matrices in inverse order to that
which the ray physically encounters the elements\cite{siegman}.

Huygens' Integral can be generalised to
an equation that includes the Ray Matrix of an overall system (even allowing
for complex ray matrix coefficients)\cite{siegman}, and which describes the entire
paraxial propagation of light through the system, including any diffraction
(but not aperture) effects. Limiting apertures (such as those forming the
polygonal mirrors of our cavity) require propagation of light to this plane,
then spatial filtering to produce the apertures.

\subsection{Polygonal Mirrors \& Nonorthogonal Basis}

Previous studies have used a nonorthogonal basis\cite{kaleidoscope} 
to describe the field
intensity across the cavity. This allows polygonal and rhombus aperture shapes
to be described exactly by the discretised mesh, whereas a standard
Cartesian discretisation would necessitate a far finer mesh to
describe the aperture with sufficient accuracy, and to not introduce
significant errors from the `jagged edges'.

This introduces complications in the numerical code, as Huygens'
integral takes on a modified Fourier form, such as the `Hankel' transform
encountered when using cylindrical coordinates\cite{siegman}.
As this causes considerable constraints in choice of aperture shapes, and is
fundamentally more difficult to visualise what is occurring, we decided to
stick with a Cartesian grid.

\section{Relevant Previous Studies}

Previous work\cite{kaleidoscope}, utilising a Fourier convolution method to
solve the Huygens' integral, have been used to investigate unstable cavities
with small Fresnel number $N_{eq}$ and linear magnifications of $1 \le M \le 2$, with
detailed study of the 8 lowest loss eigenmodes of $M=1.3$ cavities. 

In order to produce eigenmodes of fractal nature, target mirrors of regular
polyhedral \& rhomboid shape were constructed. The produced mode patterns
show self-similar structure, with higher $N_{eq}$ factors leading to a more
complex \& developed fractal structure. The fractal dimension of these
patterns were found to be $1 \le D \le 2$, embedded in the $2D$ transverse
plane of the cavity.

\subsection{Experimental Work}

A prototype laser, utilising modified Iris diagrams to produce an aperture of
variable size \& shape, has been constructed from a He-Xe
laser\cite{polygonalAperture}. Evidence was found of the excess noise factor
$K$ depending highly on aperture shape and $N_{eq}$, as would be expected
when producing fractal mode patterns, but there is little direct
experimental evidence of fractal structure in a laser beam.

\section{Numerical Methods}

Though the equations and methods used in our numerical simulation are present 
in the standard text\cite{siegman}, there
were many subtleties and nuances encountered when implementing a working
system.
Most infuriatingly, there are many different ways of formulating the
equations into a form suitable for solution on a computer - and it was
difficult to separate descriptions of the general method from specifics of
one implementation! This section along with consultation of the codes in Appendix A
 should allow one to quickly built a working code base.

\subsection{Fox-Li Power Method}

The Power method\cite{siegman} is a very simple concept from linear algebra.
When given any eigensystem, the repeated action of the operator (physical
propagation around a cavity) on an
initial eigenvector (the field distribution) will eventually lead to
convergence if the system has a dominant eigenvalue\cite{linear_algebra}. The mode found will be
the one with the highest absolute eigenvalue, corresponding to the
lowest-loss case. Convergence can be extremely slow, in particular when one
is in a region where two eigenvalues are nearly the same strength - a `mode
crossing'. Also, repeated action of the operator will cause the eigenvector 
(field profile) to tend asymptotically to either $\inf$ or $0$. Scaling of
the light intensity after every round-trip is necessary to keep the values
at some sensible (and therefore more accurate) level. 

As such, to find the lowest-loss mode pattern one simply needs to devise a
system for propagating light around a cavity and then apply repeatably until
converging at an eigenmode.

\subsection{Light Propagation}
The Huygens-Fresnel integral [Eqn. \ref{hfint}] can be described as a convolution
operation. Propagating light from one reference frame to the next is
achieved by summing contributions to the electric field from all the Huygens
wavelets. With a discretised grid of $n$ units, this requires computational
time of $O \approx n^2$ in $1D$, $O \approx n^4$ in $2D$ and quickly becomes
intractable.

\begin{figure}[htb]
\begin{center}
\includegraphics[width=9cm]{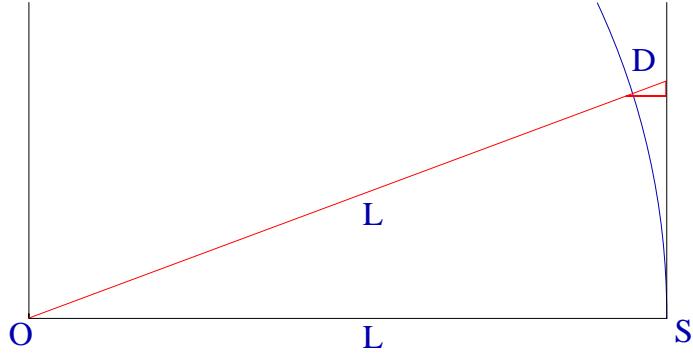}
\caption{Propagation of off axis rays introduces a phase retardation
proportional to the extra distance that the ray has to travel}
\label{chords}
\end{center}
\end{figure}

Convolution in real-space becomes a product in the Fourier
domain, which brings enormous computational savings. 
A more innate physical understanding is garnered by interpreting the
Fourier transform of a field profile as a collection of infinitely wide plane waves with
varying angle to the normal of the reference plane (Fig. \ref{chords}). 
These can then be propagated (in either direction) by making an
allowance for the different path-lengths travelled by the waves. 
This involves applying a phase change equivalent to the extra distance that
non-axial rays have to travel to reach the new plane.
The
light field distribution can then be recovered by
undertaking an inverse Fourier transform.

The path difference (Fig. \ref{chords}) is first approximated in the
paraxial approximation to be equal to the difference parallel to the plane
of the optical cavity and then the circular arc of equi-distance is
approximated as a parabola. Therefore, the extra length of a 
non-axis
plane wave is proportional to $L \lambda k^2$, with $k$ being the spectral angle. 

\begin{eqnarray}
\zeta_0(k_i,k_j) &=& \mathbb{F}(\eta_0(i,j))\\
\zeta_1(k_i,k_j) &=& \zeta_0(k_i,k_j) e^{i \pi L \lambda (k_i^2+k_j^2)}\\
\eta_1(i,j) &=& \mathbb{F}^{-1}(\zeta_1(k_i,k_j)
\end{eqnarray}

\subsubsection{Implementation}
We used the high-performance FFTW\cite{FFTW05} library routines to undertake
our Fast Fourier
Transforms (FFTs). In spite of the high efficiency, the FFT step was found
to be by far the most time-consuming step of running the simulation, which
meant that none of our code was particularly speed critical. 

Our FFT routines were found to require a rescaling to conserver
energy of
$\eta_1=\frac{\eta_0}{N}$ in $1D$ and $\eta_1=\frac{\eta_0}{N^2}$ in $2D$ after every application of
a pair of transform \& inverse transform. Like many Fast-Fourier-Transform
methods FFTW\cite{FFTW05} flips the placement of high \& 
low spectral frequencies for performance reasons. Rather than waste time 
remapping these into a scratch buffer, modulo
arithmetic was used to rearrange the array lookups in situ whilst working in
the Fourier domain.

When working on a $NxN$ discretised grid with a full-width aperture size
$A$ and $\frac{N}{2}$ locating the axis of propagation, we need to do the
following in the Fourier domain:

$      out[i][j] *= cexp (  i \pi \frac{L*\lambda}{A^2}
(((i + \frac{N}{2}) \% N - \frac{N}{2})^2 +
  ((j + \frac{N}{2}) \% N - \frac{N}{2})^2 );						      
$

\subsubsection{FFT Guard Bands}

\begin{figure}[htb]
\begin{center}
\includegraphics[width=9cm]{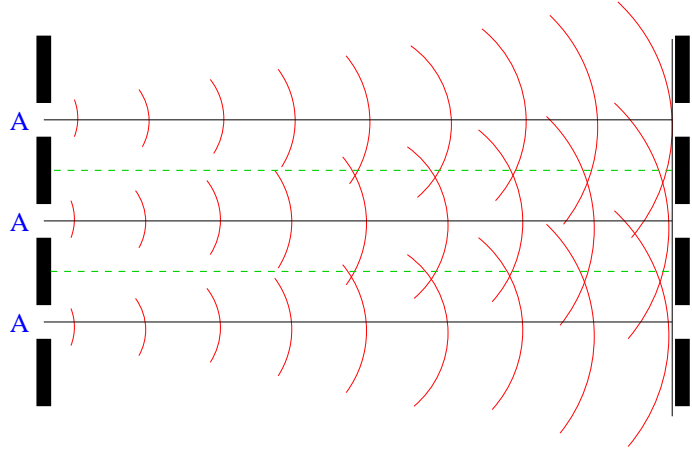}
\caption{The finite nature of the Discrete Fourier Transform results in 
an infinite series of propagating waveforms, either side of the (dotted) 
region one is using. As such, light can leak across from either side, 
producing an incorrect field distribution.}
\label{fft_bands}
\end{center}
\end{figure}

The FFT is not a true Fourier transform (which is an integral from
$-\infty$ to $+\infty$), but an infinite series of finite integrals.
Within the framework of considering the physical interpretation of the
mathematics, this can be visualised as an infinite series of screens onto
which the light is imaged. As can be seen (fig. \ref{fft_bands}) 
one needs to use 
use a Guard Band around the edges of the usefully-propagated region, in
order to stop spillover from adjoining cells.

The requirement of this $G$ factor can be indicated \cite{1975} by considering the
amount of light spilling from a single aperture uniformly illuminated
due to a combination of the overall geometric magnification and an edge
diffraction pattern.

This spillover is proportional to the equivalent Fresnel number $N_{eq}$ and 
the magnification of the setup. With esoteric cavity
designs we found that the actual required $G$ factor generally far exceeded
this amount (and was even influenced by the shape of the aperture
in $2D$).
In general we confirmed the sufficiency of the guard bands by varying them 
slightly and seeing if there was any change of the supposed eigenmode produced.

The maximum energy spillover was found 
to be less than $10^{-2}$ for reliable eigenmode formation.

The most elegant and simple test of the propagation subroutine was to
generate a test Gaussian wavefront, then propagate varying distances to
confirm that the standard beam propagation relations were reproduced.

\subsection{Lensing}
\begin{figure}[htb]
\begin{center}
\includegraphics[width=12cm]{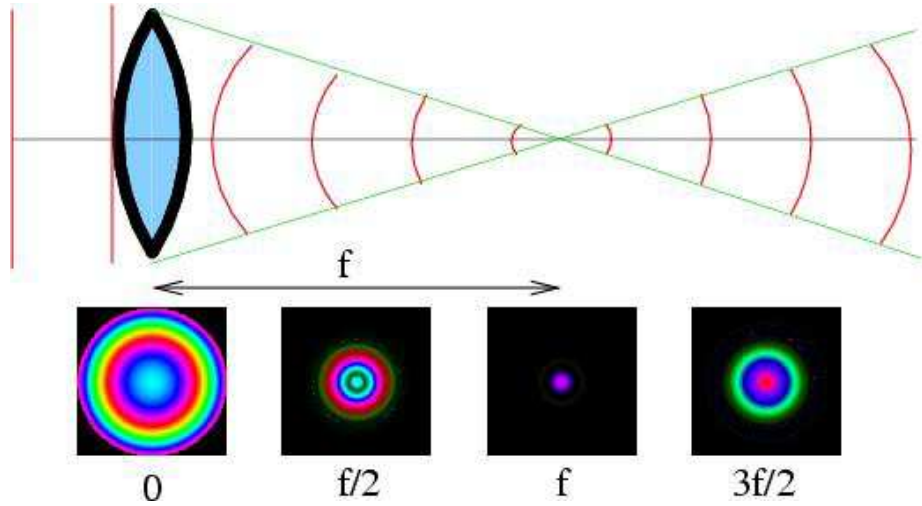}
\caption{Effect of an infinitely thin lens is to induce a spherical
curvature to the wavefront, resulting in a hitherto parallel collection of
rays converging on it. The $2D$ pictures below (colour representing phase,
intensity representing intensity) are data from our codes, which show a
circular cross-section plane wave focusing down to a diffraction limited
spot near the focal point, before expanding again.}
\label{lens}
\end{center}
\end{figure}

Our optical cavity is unfolded into an infinite series of thin lenses \&
free-space propagation.
In the paraxial approximation, with a mirror surface that causes no
overall phase-shift to the reflected light and with perfectly aligned optics, 
a lens is simulated by that of a phase shift proportional to the square of
the distance from the optical axis, scaled to produce correct focal-point
effect (fig. \ref{lens}).

Mathematically, what we are doing in the Spatial domain is:

\begin{center}
$u_1(i,j) = u_0(i,j) e^{\frac{i \pi}{f \lambda} (i^2+j^2)}$
\end{center}

On a discretised grid of $NxN$ units, with a full-width aperture of $A$,
this corresponds to pseudo-code of:

\begin{center}
$ap[i][j] *= cexp (\frac{i \pi}{f \Lambda} \frac{A^2}{N^2} ((i - \frac{N}{2})^2 +
	                               (j - \frac{N}{2})^2) );$
\end{center}

\subsubsection{Testing Correctness}

The main mechanism to test the effect of lensing was to apply a lens to 
plane-parallel light, then propagate forwards to various points (most
importantly, the focal length $f$ and the image point $2f$) to confirm that
the correct behaviour was observed. In $2D$ with a `square' chunk of light, the
focusing effect was lost amid pronounced edge diffraction, whereas a circular
beam of plane light was far more elegant (fig. \ref{lens}). 

\subsection{Bare Optical Resonator}

\begin{figure}[htb]
\begin{center}
\includegraphics[width=12cm]{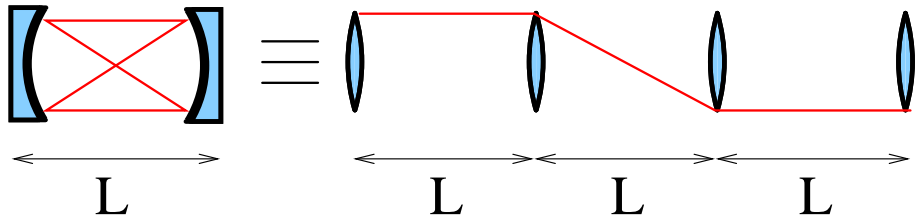}
\caption{Unfolded optical cavity.}
\label{unfolded}
\end{center}
\end{figure}

A basic bare optical resonator is one in which two mirrors face each other
across a non-interacting medium. In terms of our simulation, we
unfold these two mirrors into an infinite series of thin lenses \& propagation
over free-space (fig. \ref{unfolded}).

For a situation where the defining aperture is one of the mirrors (such as
most unstable cavities) it is possible to
condense any combination of intervening optical elements into a single ABCD
matrix. This can then be expanded into an `equivalent
lens guide'\cite{siegman} of lens \& free-space (Fig. \ref{unfolded}).
This results in a vast saving in computational time, as each ABCD element 
with a non zero B component (representing some free-space propagation) 
otherwise requires two FFTs. 
For a two-mirror cavity setup with
one mirror defining the aperture of the system this doubles the speed of
calculation.

Though there are methods\cite{siegman} of incorporating a
general ABCD matrix directly into the Huygens-Fresnel integral, we found it
better to use an equivalent Lens \& Free-Space
infinite series, as we found it far easier to debug and visualise what was 
going on. We soon garnered an intuitive grasp of how requirements of minimum
discretisation ($N$) and guard bands ($G$) varied with different cavities.

\subsection{Detection of Eigenmodes}

The Fox-Li method will eventually deliver the simulation at an eigenmode,
whereupon Eqn. \ref{eigeneqn} is held. 
It is essential to be able to detect when
an eigenmode has been reached, and derive a quantitative
figure for the eigennumber of the mode ($\gamma$).

\begin{equation}
\label{eigeneqn}
\uuline{M} \eta_n = \gamma_n \eta_n
\end{equation}

The spatial variation of the field remains unchanged after a round-trip of
the cavity - however, it is necessary for there to be an overall phase change
\& energy loss associated with the eigenmode. 
This can be represented as the $\gamma$ factor, a
complex scalar by which all the points making up the spatial representation
of the eigenmode are multiplied by to arrive at the exact pattern produced
after one more round-trip.

The unchanging nature of the gamma factor upon successive round trips, 
and the fact that all points
across the spatial domain have to simultaneously obey it when at an
eigenmode was used to simultaneously detect eigenmodes and quantify $\gamma$. 
After each
round trip of the cavity, the average $\gamma$ factor was calculated by
comparing each $\eta_1(i,j)$ with $\eta_0(i,j)$, ignoring positions
corresponding to relatively small values of $\eta$ with the justification
that the error in $\gamma$ would scale with $\frac{1}{\eta}$. Once this
average $\gamma$ figure was calculated, it was compared with all the individual
$\gamma(i,j)$ measurements. An eigenmode was detected if none of the
$\gamma(i,j)$ figures disagreed
with the overall $\gamma$ factor by more than a certain tolerance, generally
taken to be $1$ part in $10000$.

This method was found to be very flexible and established 
the most accurate possible $\gamma$ factor. 
The tolerance was derived experimentally, set
at a point where the `eigenmodes' produced no longer varied with a changing
tolerance value. It was generally found that convergence on the first
eigenmode came within $100$ passes for an unstable cavity configuration,
but exceptional cases near mode-crossings could take up to $1000$ passes.

\subsubsection{Implicit Shift}

The basic Fox-Li power method is only capable of finding the lowest loss eigenmode
for the resonator. In order to allow higher-loss modes to be found, one
must somehow cancel the presence of the dominant lower modes.

This can be accomplished by using knowledge of the $\gamma$ factor
associated with a given mode to destroy the contribution of that mode in the
overall pattern. Considering the measured light field as a summation of
various eigenmodes, then after one round trip we will 
a field of these modes multiplied by their respective $\gamma$ factors:

\begin{eqnarray}
\eta_0 &=& u_1 + u_2 + u_3 + u_4 \ldots \\
\eta_1 &=& \gamma_1 u_1 + \gamma_2 u_2 + \gamma_3 u_3 + \gamma_4 u_4 \ldots
\end{eqnarray}

If we take our round-trip profile $\eta_1$ and subtract a known
 $\gamma_1\eta_0$ factor (using the previously found $\gamma$ and the prior
light field distribution), then we will destroy
any presence of the this eigenmode in $\eta_1$. However, since this step
can never be perfect (due to inaccuracy in our estimation of $\gamma$) and
the natural tendency of the system to want to generate the lowest-loss
eigenmode - we must apply this cancellation repeatably during the simulation.
Each mode that we
find gives us another $\gamma$ factor that we can apply in succession in order
to find higher modes. This is equivalent to the Shift method in linear
algebra\cite{linear_algebra}.

The most effective method of mode-searching that we found was to iterate in
successive round-trips of the cavity over a `carousal' of the found $\gamma$ 
factors, then
allow a number of normal unperturbed round trips (generally $10$) to see 
if a new eigenmode would
stabilise before already-known lower-loss eigenmodes arose from the background
noise. For instance, if we consider there to be an error of $\sigma$
associated with each $\gamma$ factor, and a carousal of three known
eigenmodes the process will be:

\begin{eqnarray}
 \eta_0 &=& u_1 + u_2 + u_3 + u_4 \ldots \\
 \eta_1 &=& \sigma_1 u_1 + \gamma_2 u_2 + \gamma_3 u_3 + \gamma_4 u_4 \ldots\\ 
 \eta_2 &=& \sigma_1 \gamma_1^2 u_1 + \sigma_2 u_2 + \gamma_3^2 u_3 + \gamma_4^2 u_4 \ldots\\ 
 \eta_3 &=& \sigma_1 \gamma_1^3 u_1 + \sigma_2 \gamma_2 u_2 + \sigma_3 u_3 + \gamma_4^3 u_4 \ldots \\ 
 \eta_4 &=& \sigma_1 \gamma_1^4 u_1 + \sigma_2 \gamma_2^2 u_2 + \sigma_3 \gamma_3 u_3 + \gamma_4^4 u_4 \ldots
\end{eqnarray}

Clearly - new modes stop being found once the error in $\sigma$ is such that
the higher-loss higher-order mode is never able to dominate (i.e. when
$\sigma_1 \gamma_1^{n-1} \approx \gamma_i^n$). Modes often
require a number of such cycles to become dominate \& be identified but
even imperfect mode cancellation allows the $u_n$
part of the overall $\eta$ distribution to be sufficiently strong so that it
stabilises to an eigenmode.

\subsection{Polygon Apertures}

A method was sought to be able to easily produce regular polygon apertures.
The best method found was to use a standard point inclusion in polygon test
\cite{pnpoly}, then fill in an aperture mask using this selection criteria.

\subsection{Outputs}

A simple data format was used containing tab-separated values for the
various measurements across the middle of the aperture (or along the
infinitely thin slice when running in $1D$). These values included the
location $x$, complex and imaginary parts of $\eta$, the intensity ($I\alpha
\eta^2$) and phase ($\phi_x=atan\frac{\Im(\eta_x)}{\Re(\eta_x)}$).

The $\eta$ field
variation across the centre of a $2D$ square aperture should agree exactly 
with a $1D$ infinitely thin slit with identical cavity configuration. This 
was used as an indicator that the overall $2D$ functioning was correct. 

\subsubsection{Representing Phase \& Intensity}

In order to show both phase and intensity information in one figure, a
routine was written to output colour representations of the transverse
light field, with the `brightness' of the pixel representing the
intensity, and the phase ($0 \to 2\pi$) sampling a colour from a standard $HSV$
colour wheel\cite{hsv}, with no phase ($\phi=0$) as cyan. These could then easily
be separated at a later date into monochrome intensity plots (desaturation) 
or pure phase plots (full saturation).

\subsection{Sampling Fractal Dimension}

There are a number of, seemingly contradictory, methods of defining the
fractal dimension of a particular object ($D$). As we are dealing with
graphs (the transverse eigenmodes) with an unclear relationship between
the scaling of the physical dimension versus the intensity, simple box
counting methods are less than ideal. A method of calculating the Hurst
exponent using the rescaled range was found to be the easiest to implement
and was implemented in our codes.

The Hurst exponent can be considered a measurement of the persistence of the
function - the extent to which it is predictable. Fractal dimension, for a
$2D$ graph is related by:

\begin{equation}
D = 2-H
\end{equation}

A predictable linear varying function will have
a Hurst exponent of 1 (range scales directly with sample size) and therefore a
fractal dimension of 1 - it is a $1D$ object embedded in a $2D$ graph.
Conversely, the more complexity in the mode pattern, the lower the Hurst
exponent and the higher fractal dimension. However - pure noise has a Hurst
coefficient of zero, implying the highest possibly degree of fractal structure.
Therefore, though Hurst analysis can be used to quantify the degree of
fractal structure, it is not sufficient to declare whether a particular
structure is a (self-similar) fractal or not.


\section{Results}

\subsection{Equivalent Confocal Cavity}

A general cavity can be converted into an equivalent confocal form by
placing the $ABCD$ matrix between two opposite lenses. This can then be
easily unfolded into an equivalent lens guide,
allowing one to halve the computational effort necessary to locate
eigenmodes. We generally investigated symmetric cavities ($g_1=g_2$).

Eigenvalues are located on a convex hull in the complex plain, 
but due to the way that the
carousal-algorithm `hops' shifts the origin on the plane, the modes are not
discovered in strict order of lowest loss but that each found eigenmode is
the one furthest away (therefore greatest in magnitude) 
from the $\gamma$ values being used to suppress the lower order eigenmodes. 

\begin{figure}[htb]
\begin{center}
\includegraphics[height=6cm,angle=270]{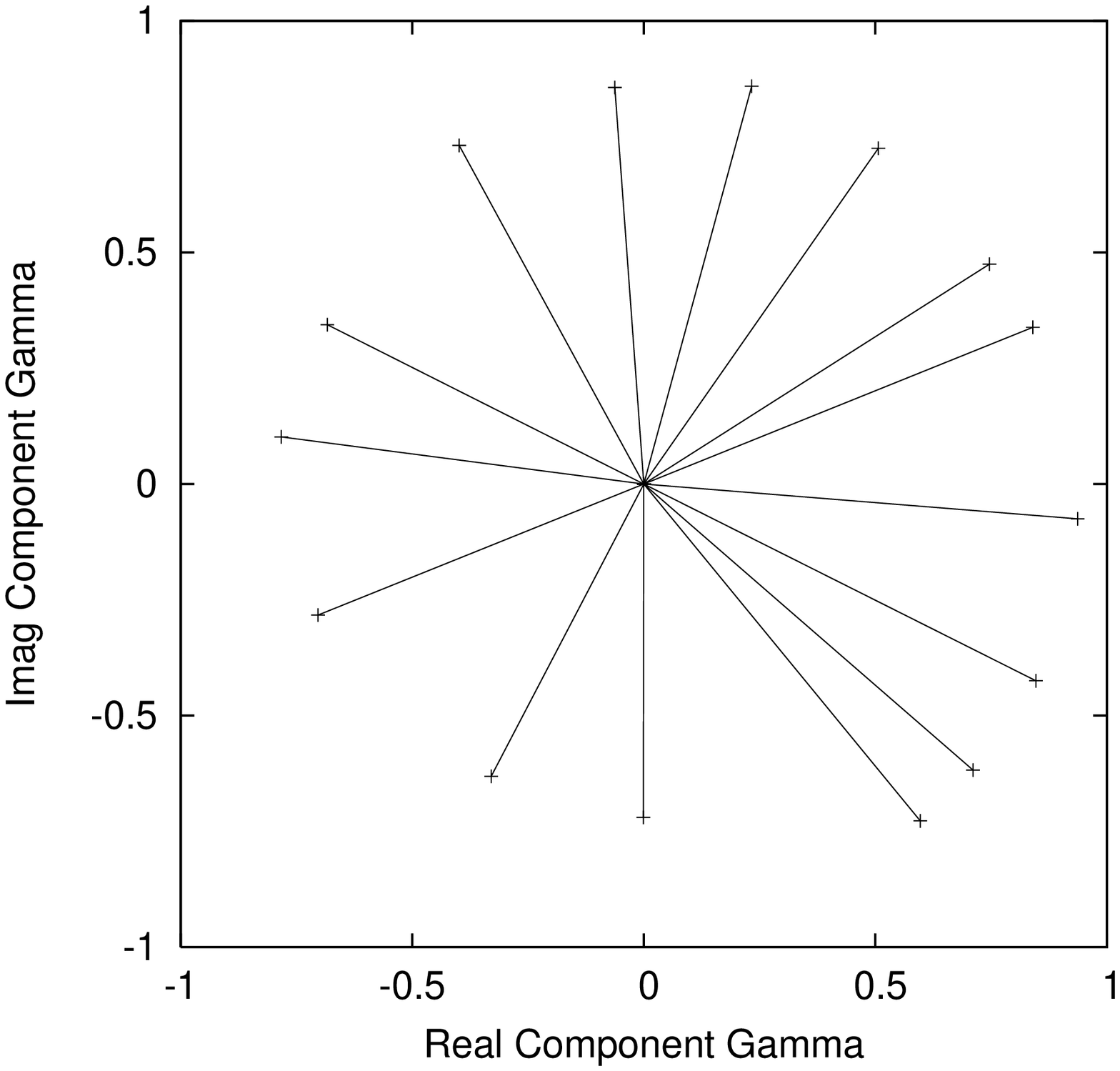}
\includegraphics[height=6cm,angle=270]{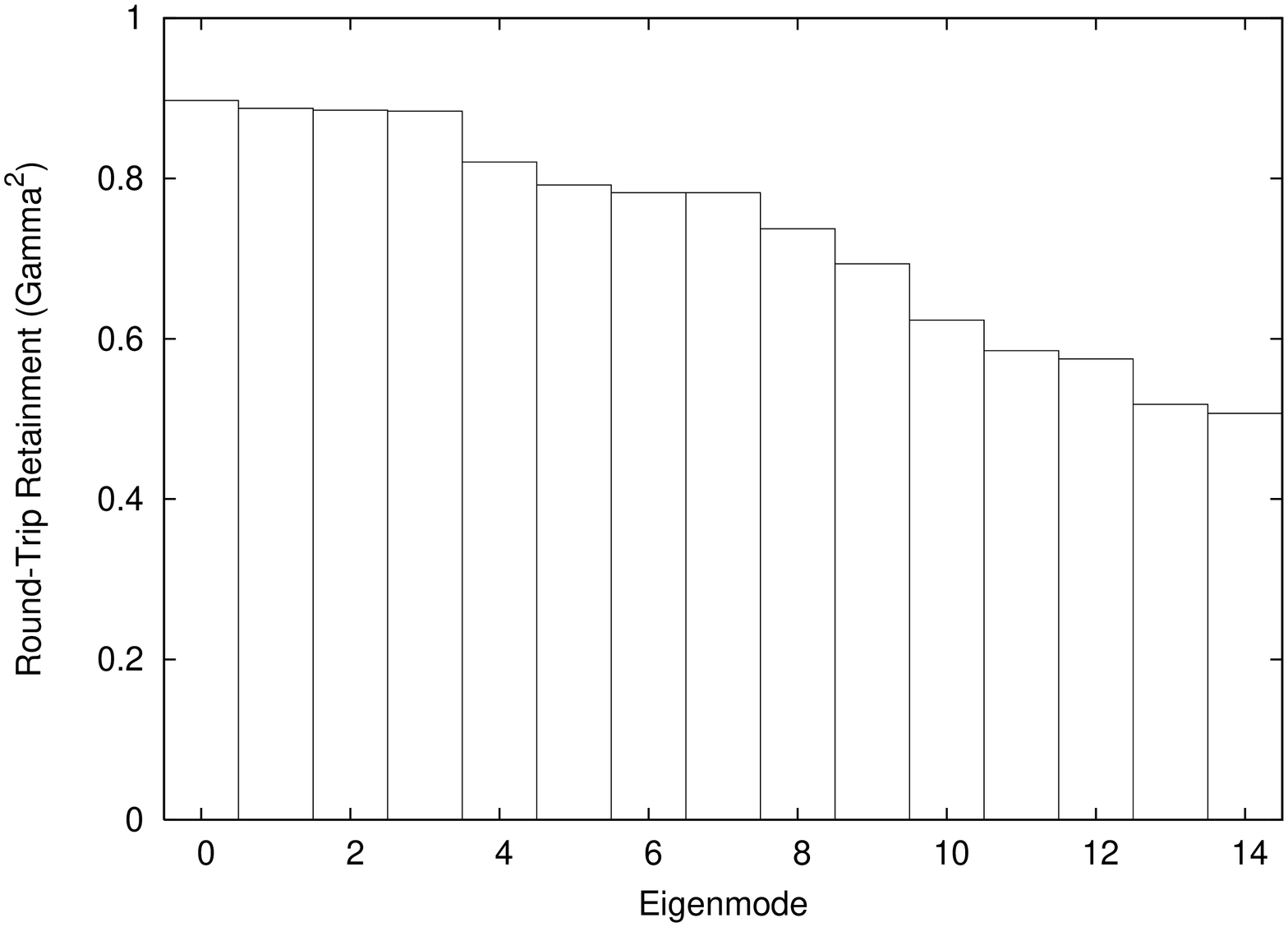}
\caption{Gamma factors for the first 15 Eigenmodes of a low-magnification
symmetric cavity ($M=1.1$) with a hexagonal limiting aperture. 
Each `spoke' of the wheel represents the gamma
factor of the eigenmode, the square of the length of which (the absolute value of the
complex $\gamma$ factor) represents the energy retained after each round trip
in this eigenmode.}
\label{eigens}
\end{center}
\end{figure}

\begin{figure}[htb]
\begin{center}
\includegraphics[width=13cm]{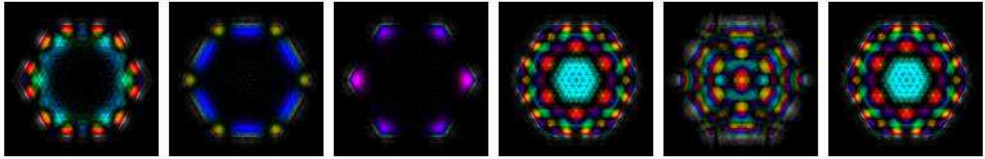}
\caption{
Intensity plots of the 6 lowest loss eigenmodes from (fig. \ref{eigens});
colour is phase of light on HSV\cite{hsv} colour wheel. 
}
\label{lowest_6}
\end{center}
\end{figure}

A typical set of eigenvalues is represented (fig. \ref{eigens}),
along with the associated $2D$ plots (fig. \ref{lowest_6}). Weak self similar
structures are demonstrated in these transverse patterns, with the structure
of the aperture (an n-side polygon) being replicated at smaller and smaller
scales.

\subsection{Confirmation of Codes Accuracy}

In order to assure that our codes were finding eigenmodes correctly, we
replicated data from the literature, in particular Fig. 4
of \cite{confirmation}).
This consisted of locating lowest-loss eigenmodes for a symmetric cavity
with magnifications $1.8$ \& $1.9$, using a relatively low equivalent Fresnel number of
$49.4$. Agreement is extremely good, and gave us
confidence in the ability of our codes to find accurate eigenmodes.

\begin{figure}[htb]
\begin{center}
\includegraphics[height=6cm,angle=270]{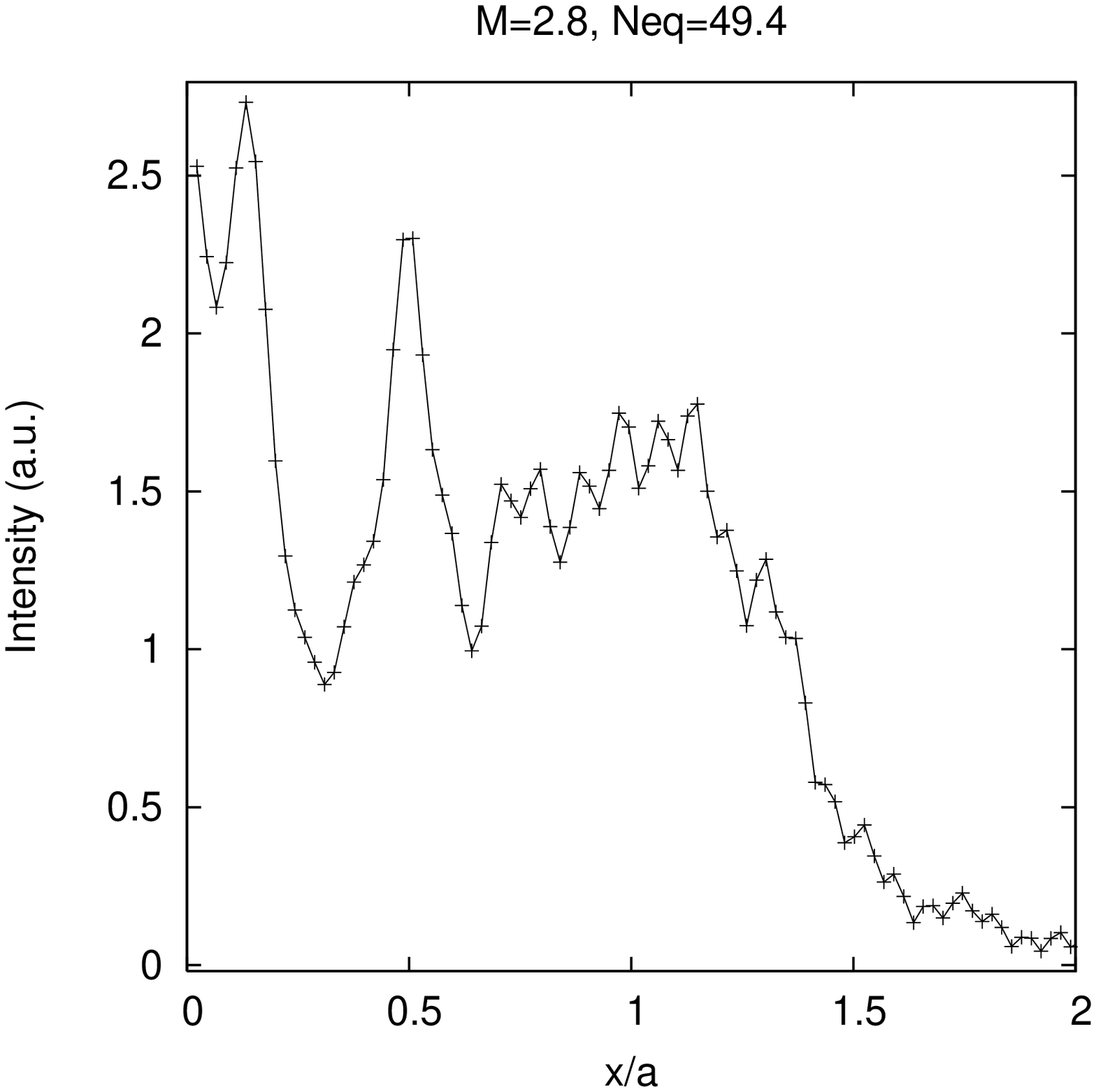}
\includegraphics[height=6cm,angle=270]{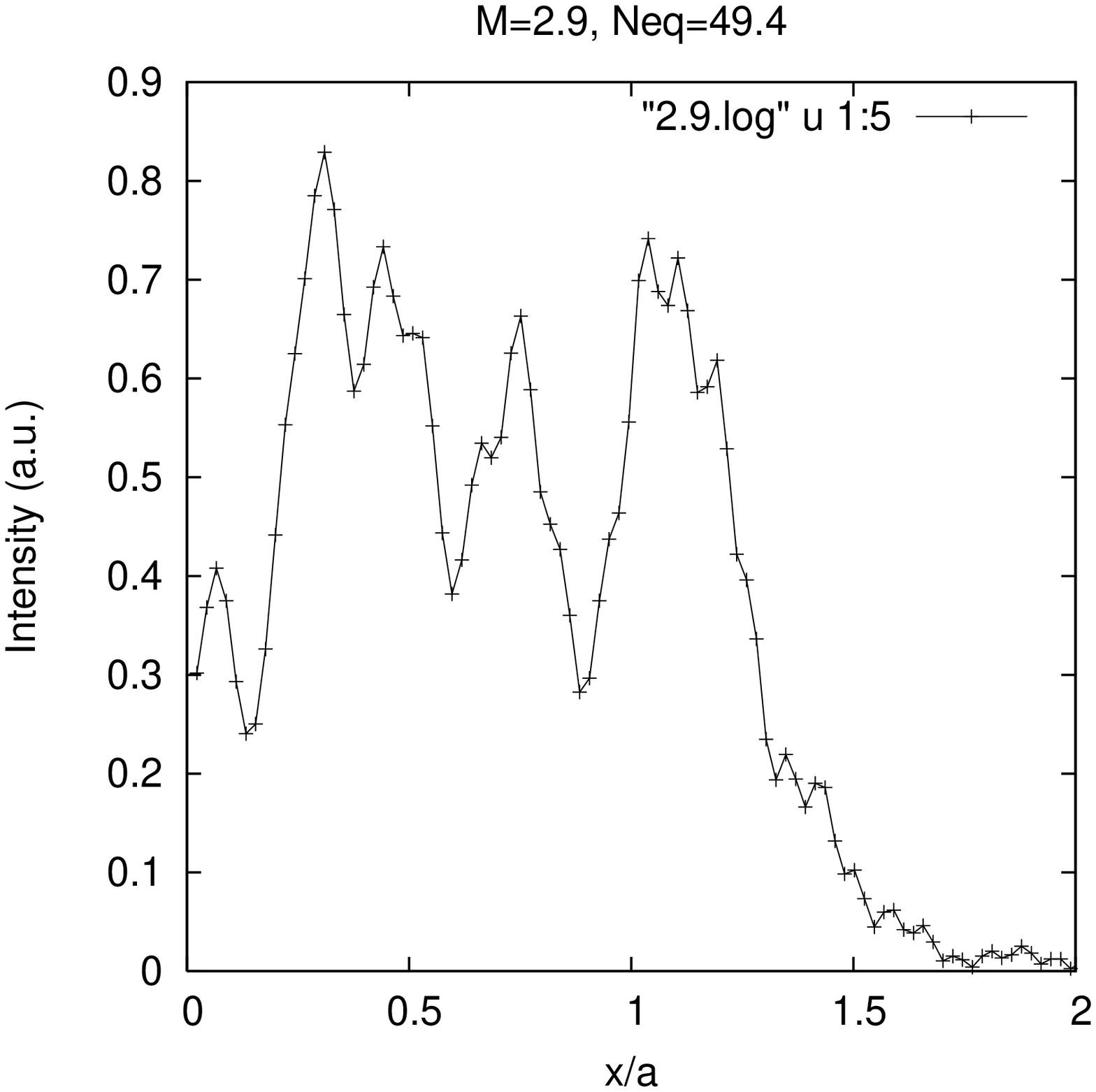}
\caption{
Intensity mode-profiles for a $1D$ (and similarly, cross section on square
aperture $2D$) unstable confocal cavity with $M=2.8$,$2.9$. $N_{eq}=49.4$
}
\label{2.8_graph}
\end{center}
\end{figure}

\begin{figure}[htb]
\begin{center}
\includegraphics[width=6cm]{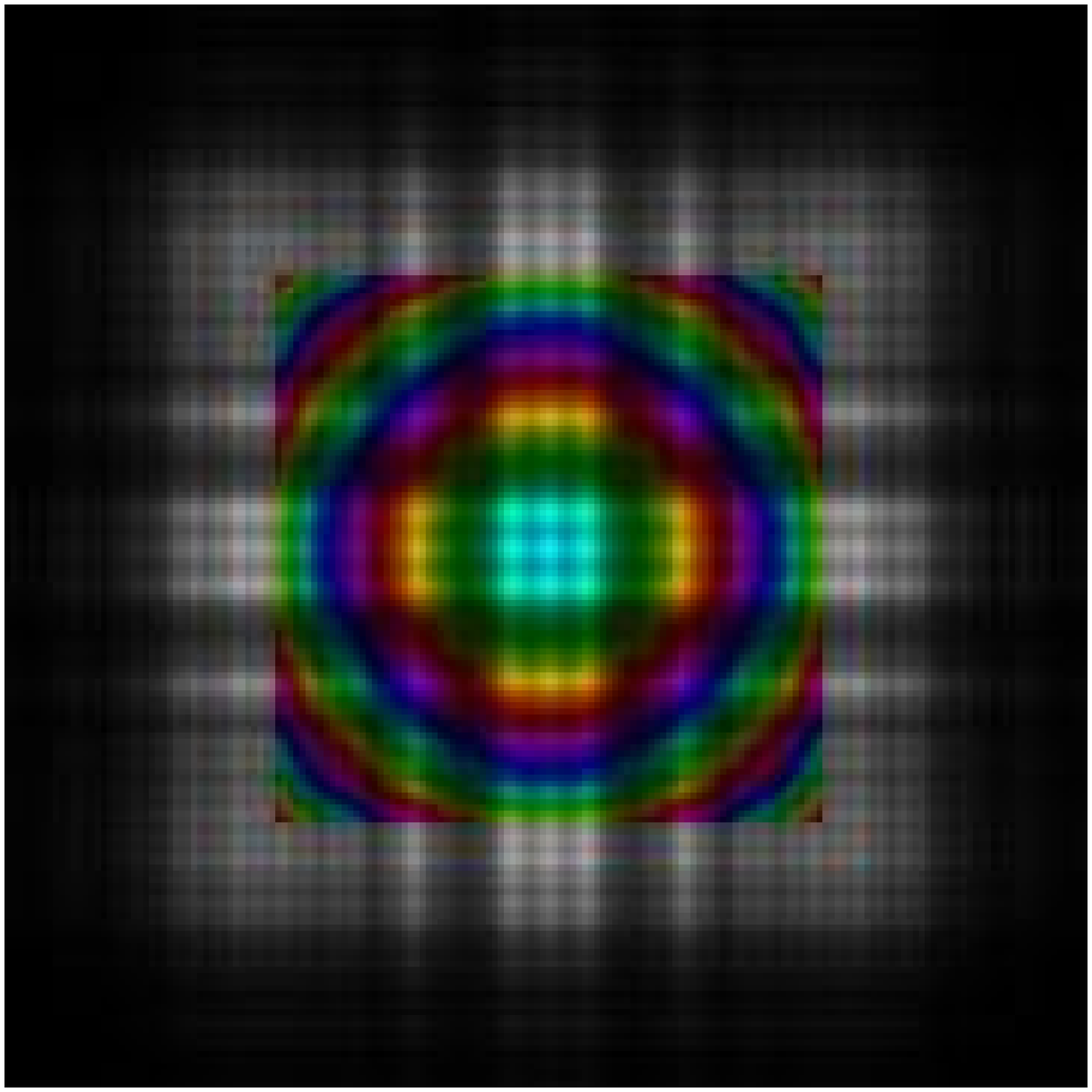}
\includegraphics[width=6cm]{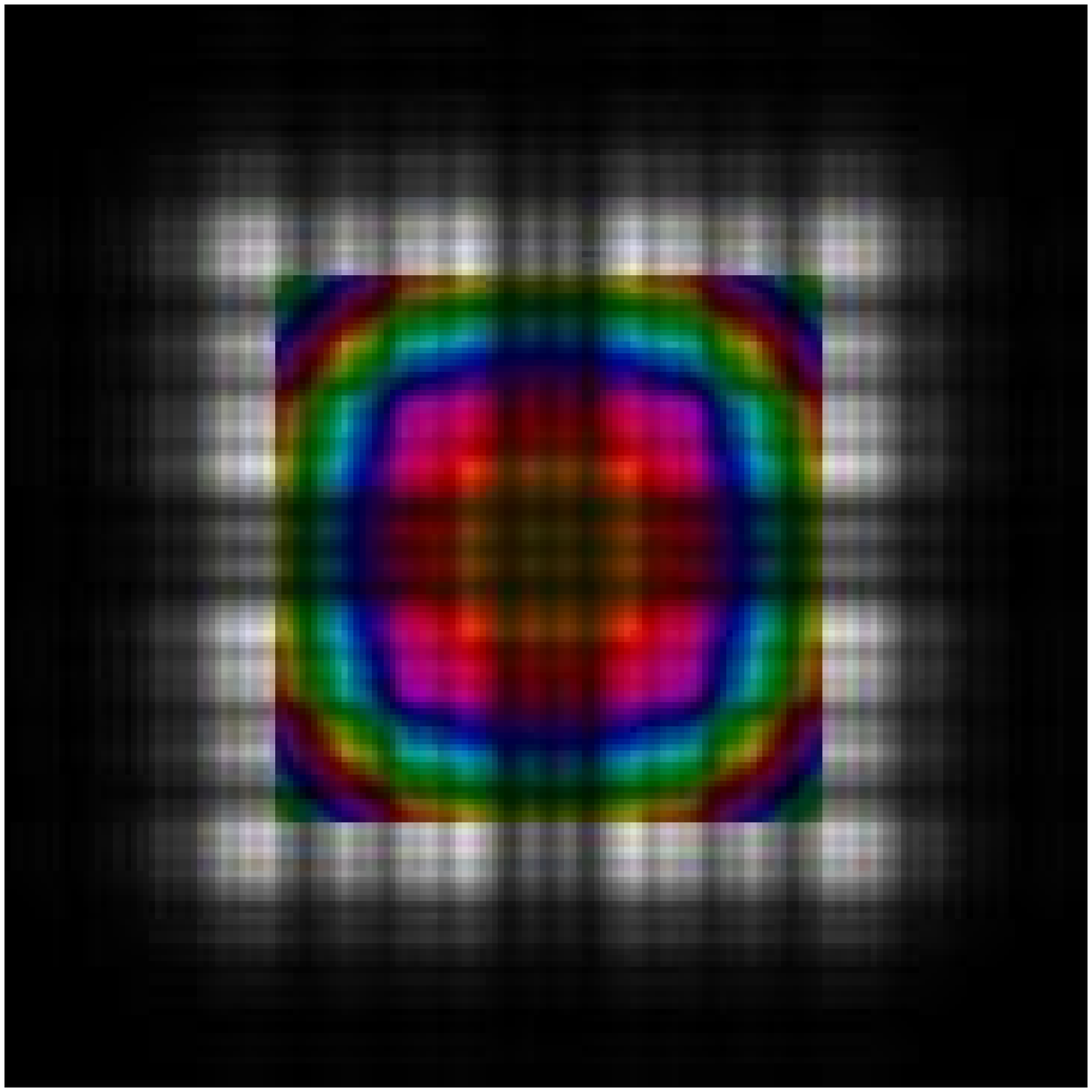}
\caption{
Corresponding transverse $2D$ plots.
}
\label{2.8_cropped}
\end{center}
\end{figure}

\subsection{The Conjugate Plane}

\begin{figure}[htb]
\begin{center}
\includegraphics[width=7cm]{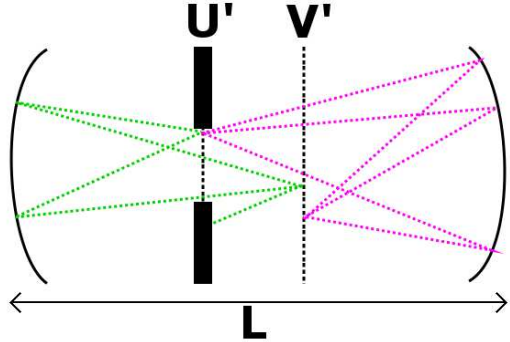}
\caption{
Location of the $U'$ and $V'$ conjugate planes in a confocal cavity, by a
simple ray tracing argument.
}
\label{conjugate_planes}
\end{center}
\end{figure}

Considering a symmetric cavity, with $g<-1$, one can identify by a simple
ray tracing argument the existence of two conjugate planes. These planes $U'$
\& $V'$ image onto each other after a half-trip of the cavity (with a
magnification of the half-trace of the cavity: $m=\sqrt{M}$), and onto
themselves after a full round trip of the cavity (Fig. \ref{conjugate_planes}).

\begin{figure}[htb]
\begin{center}
\includegraphics[width=9cm]{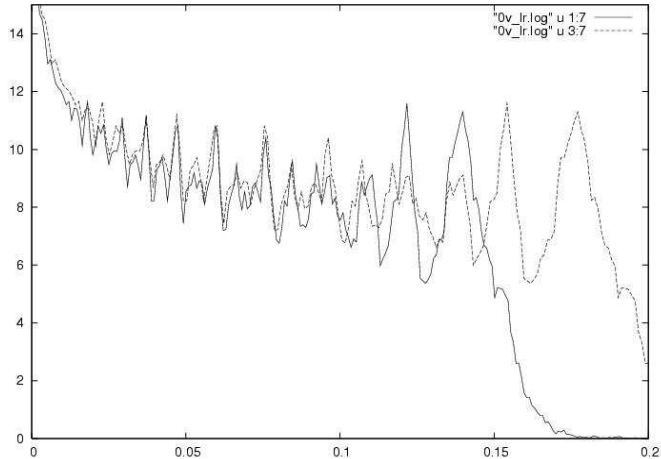}
\caption{
Eigenmode projected to $U'$ conjugate plane, plotted against a $M$
stretched version of itself - showing clear self similarity. $N_{eq}=49.4$,
$M=1.3$.
}
\label{fractal_conj}
\end{center}
\end{figure}

With the simple case of where the defining aperture of the cavity is at one
of the mirrors, and then propagating to a suitable conjugate plane, one
discovers that the pattern of the eigenmodes bare a very clear fractal character. 
In fact, if one overlaps the mode pattern against a Magnification-factor
stretched version of itself (Fig. \ref{fractal_conj}), there is extremely
good agreement in pattern (peaks and troughs) but not the magnitude of the
eigenmode.

Perhaps this should not be too surprising - by its very definition, the $U'$
and $V$ planes are self-imaging, but do so with a round trip magnification.
Any eigenmode rendered at the image plane will have to obey such a
situation, and so must display a self-similar nature.

Generating these modes was extremely computationally demanding, the cavity
setup requiring an extremely large FFT Guard Band, far in excess of what is
predicted by considering the $N_{eq}$ \& $M$ parameters. This is believed to
be due to the naive (non ABCD equivalent lens guide) stepwise method by
which we propagated light around these more complex cavities.

\begin{figure}[htb]
\begin{center}
\includegraphics[height=9cm]{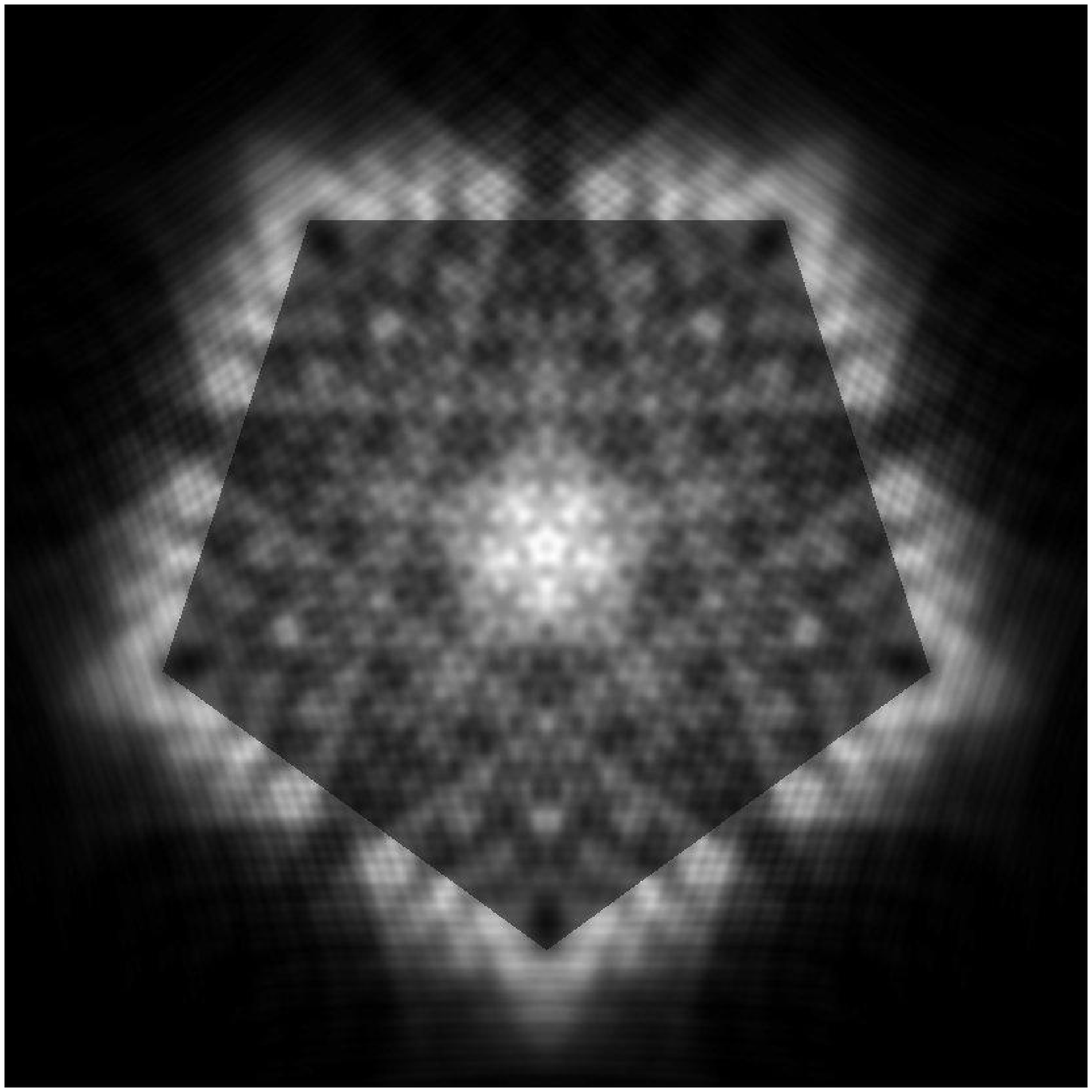}
\includegraphics[height=9cm]{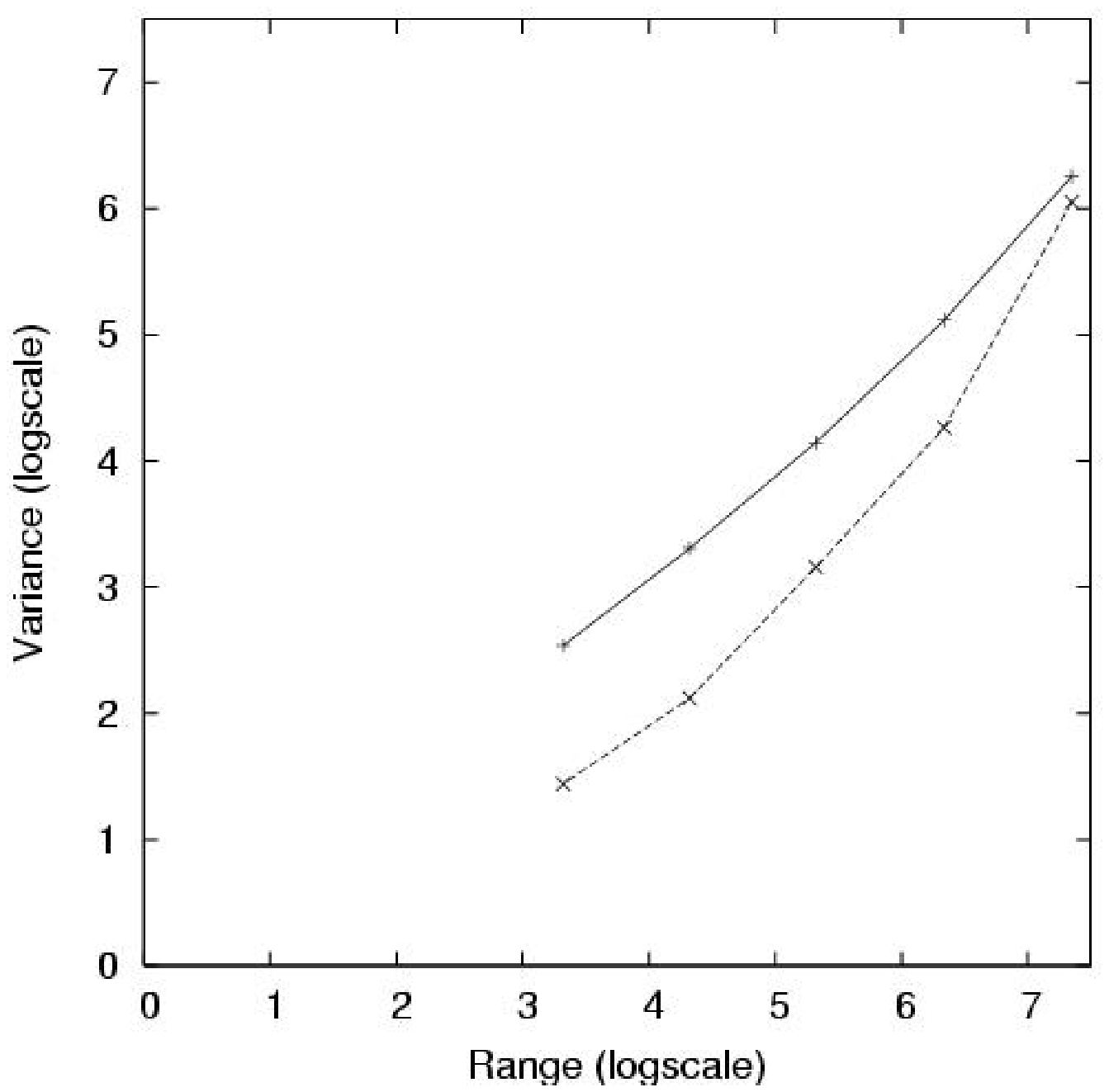}
\caption{
$2D$ Intensity plot and Hurst analysis of first eigenmode at conjugate plane with
$M=1.3$, $N_{eq}=49.4$. The Pentagonal aperture has been slightly toned to
allow its location; that pattern does not really exist in the eigenmode.
}
\label{conjugate_2d}
\end{center}
\end{figure}

As such this was even slower in $2D$ but we did manage to capture a
lowest-loss eigenmode\ref{conjugate_2d}. Curiously, there was no
pattern observable in the phase information whatsoever - just uniform
circular fringes. The reasons for this are not well understood.

The Hurst analysis for a $1D$ eigenmode (Fig. \ref{conjugate_2d}) 
with identical cavity parameters, at
both the aperture defining mirror (shallow line) and conjugate plane
(steeper line) shows a clear difference between Hurst coefficients. The
eigenmode rendered at the conjugate plane has a fractal dimension of $D=1.977$, 
whereas at the aperture mirror $D=1.84$. The conjugate plane possesses a far higher degree
of fractal structure, which confirms the qualitative appreciation of the `stretched
fit' (Fig. \ref{fractal_conj}).

\subsection{Moving the Location of the Aperture}

The fractal eigenmode produced at the conjugate plane is not perfect. It was
hypothesised that this was due to the defining aperture of the system occurring at a
non-self-imaging location, and that the quality of the fractal would be
increased by moving the limiting aperture to the imaging plane.

However, this was found to be a numerically impossible situation to
simulated - as the $ABCD$ matrix evaluated for the round trip between the
image plane and itself has a zero B value. This corresponds to there being
no effective distance between the two planes, and an infinite equivalent
Fresnel number.

In order to avoid this numerical impossibility, we attempted to shift the
aperture towards the complex pattern, and see whether an improvement in the
fractal fit of the eigenmode occurred. We produced some qualitative evidence
for this increase in fractal quality as the aperture shifted towards the
conjugate planes, but were limited from quantifying it by an inability to define
mathematically how good the fractal fit actually was, especially as
$N_{eq}\to+\infty$ and so the amount of patterning increased.

\section {Video Feedback}
A number of articles\cite{monitor}\cite{pixellated_video_feedback} draw a
parallel between the fractals character of unstable resonator modes, and the
patterns produced by a `Monitor-Outside-a-Monitor' pixelated video
feedback. The process of geometric magnification in the two systems is the
same, but instead of having the Huygens-Fresnel (Eqn. \ref{hfint}) to select
which part of the field contributes to the next intensity distribution (i.e.
the vector sum of the Cornu spiral), a pixel-function is defined that allows for 
the overlap \& finite size of pixels.

\begin{figure}[htb]
\begin{center}
\includegraphics[width=12cm]{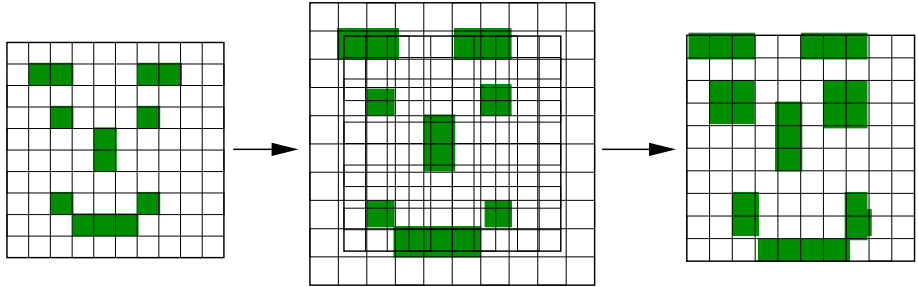}
\caption{Schematic of Video-Feedback.}
\label{vf_fig}
\end{center}
\end{figure}

As shown in Fig. \ref{vf_fig}, the `pixel-function' defines the way in which
the magnified pixels of one frame influence the pixels produced in the
next frame. 
Each successive frame is comprised of a magnification of the previous
pattern, with distorted edge effects (caused by the mismatched grid
overlap) that have a clear parallel with edge diffraction ripples occurring
in a cavity simulation.

With a simple `perfect' grid pixel function, no fractal behaviour is
demonstrated. Real video fractals occur due to the pixel structure
(the finite size of the photo-sensitive element in the CCD grid) causing an
uneven overlay of the grid.
The perfect grid situation is equivalent to the geometric limit of 
the cavity simulation - where
no diffraction effects take place and no patterned eigenmode exists.

\begin{figure}[htb]
\begin{center}
\includegraphics[width=6cm]{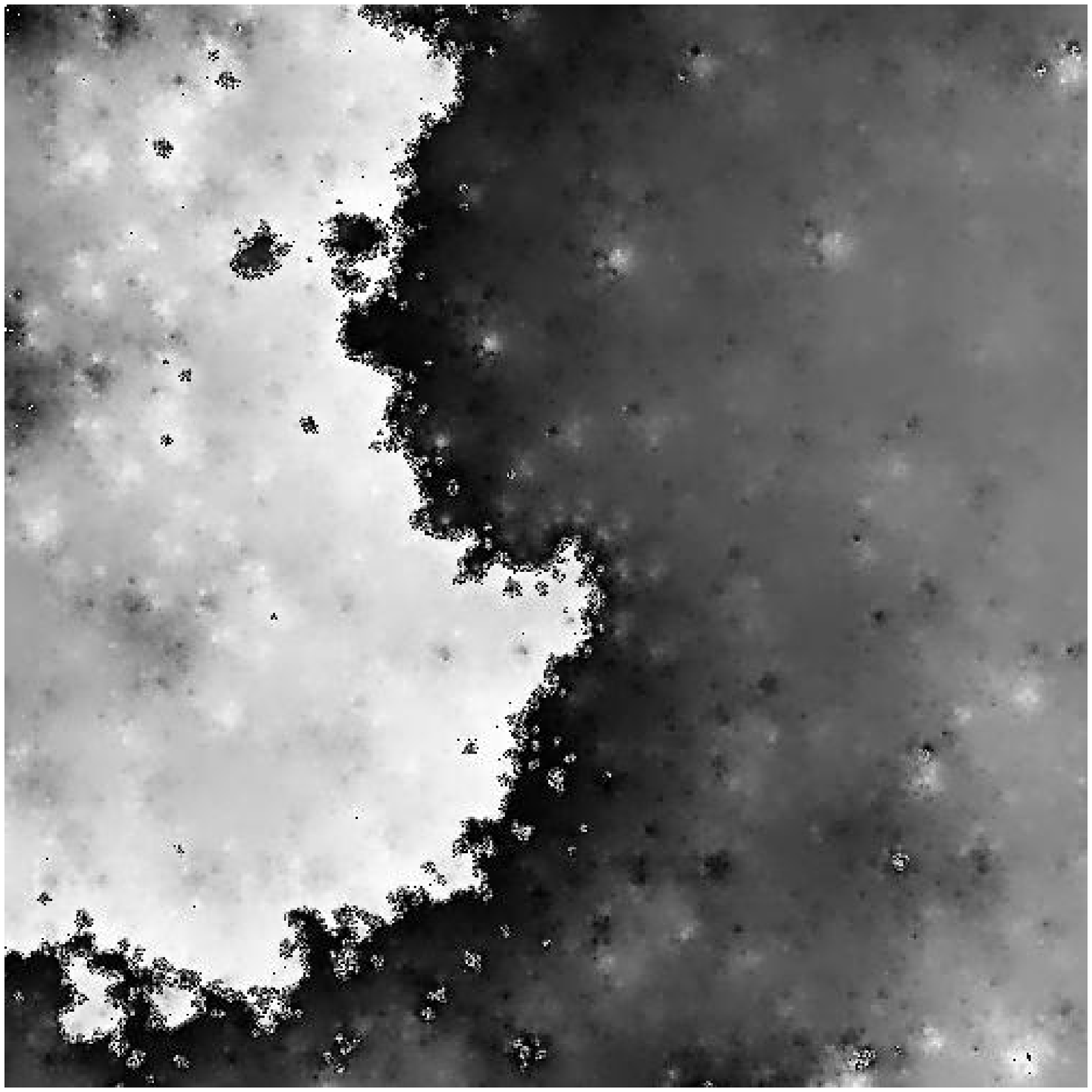}
\includegraphics[width=6cm]{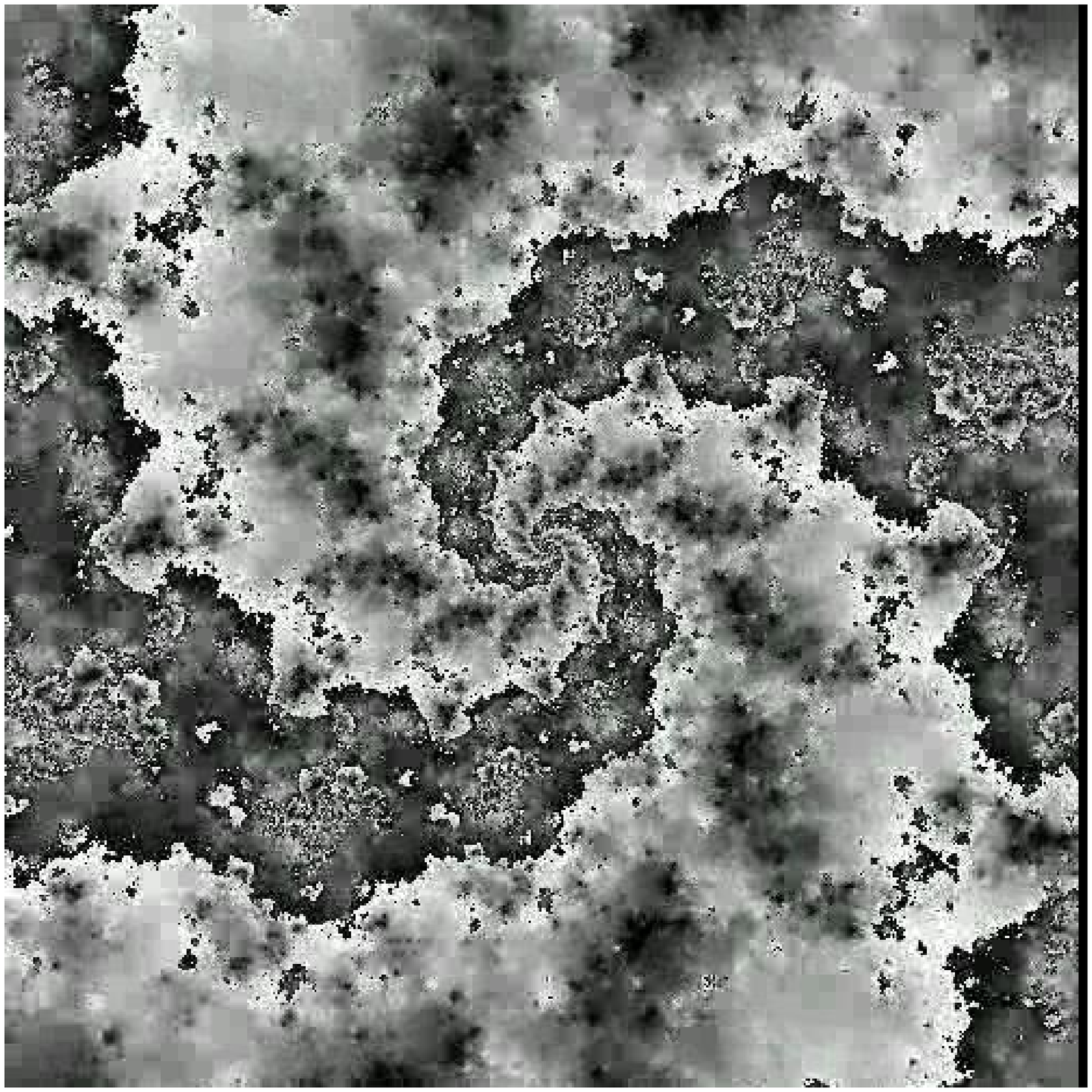}
\caption{Typical video-feedback `eigenmode' - imprecise fractal
self-similarity expressed on the `curved coastline'. $M=1.4$,
$p_{width}=0.65$ \& $M=1.3$, $p_{width}=0.6$ respectively.}
\label{video_eigenmode}
\end{center}
\end{figure}

There wasn't enough time for the codes written simulating video-feedback to
acquire sufficient finesse (in particular with the definition of the pixel
function) to be able to make direct quantitative predictions of real cavity 
eigenmodes such as has been demonstrated in \cite{monitor}
However very similar behaviour was observed in both simulated video feedback
\& the laser cavity simulation. In
particular, first order eigenmodes (Fig. \ref{video_eigenmode}) with fractal character are 
readily established by running light around the feedback system (i.e. in a
parallel to the Fox-Li power method), which are entirely indifferent to 
starting conditions, suggesting that they result from the underlying 
setup of the feedback system rather than 
the initial field distribution.

\section{Conclusions}

A set of flexible and efficient codes were produced that could accurate
predict eigenmodes produced in an arbitrary \& easily-specified bare optical
resonator  
configuration using the Fox-Li power method. 
Higher order modes could be located (at computational
expense) by the shift-method. The codes can seamlessly change between $1D$
and $2D$ cavities with $n-sided$ regular polygon mirrors.

The vast majority of time in the project was taken up in developing
these codes and proving them qualitatively against the literature, 
leaving little time to actually use them to explore the behaviour of
unstable resonator modes.

However, we clearly demonstrated that some unstable modes have self-similar
characteristics, that an n-sided polygon aperture has that motif repeated at
smaller and smaller sizes in the eigenmode and that a far higher quality of
fractal eigenmode is rendered by projecting to an image plane in a
symmetric $g<-1$ resonator.

Further work could use these codes to study the conjugate plane in greater
detail, and attempt to derive would would occur in a real system with the
aperture creating an infinite equivalent Fresnel number situation. The
biggest stumbling block that we found to this was to find a suitably
rigorous way of defining how accurate the M-stretched pattern (Fig.
\ref{fractal_conj}) matched itself. We investigated using a least-squares
method to compare, but due to the lack of magnitude conservation in the
stretched pattern, we found it would be necessary to use some kind of
rescaled range at the very least. Alternatively, one could break down the
pattern into a simplified description - such as locating the peaks \&
troughs, then comparing these; or using a spectral (FFT) method to
investigate self-similarity in the bandwidth occupied by the pattern. There
is also the unresolved issue of considering how to counteract the effect of
the increasing Fresnel number in the calculations.

These codes could also be used to investigate what happens at the stability
boundaries on the g-factor diagram, investigating whether the transition
from stable to unstable modes is as truly sharp as suggested in the
reference texts\cite{siegman}.

Codes were developed to simulate a basic Video-Feedback system, and the
fractal nature \& ability of a system to develop eigenmodes independent of
the starting field distributions was demonstrated. Further work could
develop these into a (far faster) method of predicting eigenmodes for an
optical cavity\cite{monitor}, and could even be used to `seed' a standard
Fox-Li power method near an eigenmode such as is commonly done to improve
accuracy of Prony \& VS eigenmodes.

\clearpage
\bibliography{report}

\begin{thebibliography}{10}

\bibitem{monitor}
Jogannes Courtial and Miles~J. Padgett.
\newblock Monitor-outside-a-monitor effect and self-similar fractal structure
  in the eigenmodes of unstable optical resonators.
\newblock {\em Physical Review Letters}, 85(25):5320--5323, 2000.

\bibitem{polygonalAperture}
G.~P.~Karman et~al.
\newblock Excess-noise dependence on intracavity aperture shape.
\newblock {\em Applied Optics}, 38:6874--6878, 1999.

\bibitem{pixellated_video_feedback}
Jonathan~Leach et~al.
\newblock Fractals in pixellated video feedback.
\newblock {\em Contemporary Physics}, 44(2):137--143, 2003.

\bibitem{kaleidoscope}
McDonald et~al.
\newblock Kaleidoscope laser.
\newblock {\em J. Opt. Soc. Am.}, 17(4):524--529, 2000.

\bibitem{pnpoly}
W.~Randolph Franklin.
\newblock Point inclusion in polygon test.
\newblock
  \url{http://www.ecse.rpi.edu/Homepages/wrf/Research/Short_Notes/pnpoly.html},
  1970.

\bibitem{FFTW05}
Matteo Frigo and Steven~G. Johnson.
\newblock The design and implementation of {FFTW3}.
\newblock {\em Proceedings of the IEEE}, 93(2):216--231, 2005.
\newblock special issue on "Program Generation,Optimization, and Platform
  Adaptation".

\bibitem{confirmation}
G.S.~McDonald G.H.C.~New, J.P.~Woerdman.
\newblock Excess noise in low fresnel number unstable resonators.
\newblock {\em Optics Communications}, 164:285--295, 1999.

\bibitem{linear_algebra}
Stanley~I. Grossman.
\newblock {\em Elementary Linear Algebra}.
\newblock Wadsworth Publishing Company, 1984.

\bibitem{siegman}
Siegman.
\newblock {\em Lasers}.
\newblock University Science Books, 1986.

\bibitem{pLasers}
Orazio Svelto.
\newblock {\em Principles of Lasers}.
\newblock Plenum, 1989.

\bibitem{1975}
Edward~A. Sziklas and A.~E. Siegman.
\newblock Mode calculations in unstable resonators with flowing saturable gain.
  2: Fast fourier transform method.
\newblock {\em Applied Optics}, 14:1873--1889, 1975.

\bibitem{hsv}
Wikipedia.
\newblock Hsv color space --- wikipedia{,} the free encyclopedia.
\newblock
  \url{http://en.wikipedia.org/w/index.php?title=HSV_color_space&oldid=5028301%
7}, 2006.
\newblock [Online; accessed 29-April-2006].

\end{thebibliography}
\bibliographystyle{plain}
\clearpage

\appendix

\section{Cavity Simulator Codes}
\lstset{basicstyle=\scriptsize,breaklines=true}
\begin{lstlisting}
/* Cavity Sim
 * MSci Physics Project QOLS07, Imperial College 
 * Jarvist Frost & Benjamin Hall 2005-2006
 */

#include <complex.h>
#include <fftw3.h>
#include <math.h>
#include <stdio.h>
#include <stdlib.h>
#include <time.h>

#include "pnpoly.c"		//points in a polygon detection code

//#define TWO_DIMENSIONAL

#define N (1024*16)
#define G (0.2) 	//guard band - expressed as fraction of grid we SHOULD uss
#define TOLERANCE 0.0001	//tolerance of points in testing for eigenmode
#define SAMPLEPOINTS 200	//number of random points to sample to test for Eigenmode
#define MAX_EIGENS 10

#define SPACERS 10 //number of non-shift applying 'spacers' in eigenvalue meth

#define A 1.414E-2			//was 1mm
double L = 1.0; //3.14;		//was 0.001
double FOCAL = -100000;
double LAMBDA= 10e-08; //1.0E-7; //0.4488E-6; //0.11225E-6; //0.069754E-6; //0.067349E-6; //0.069754E-6 //1 micron

int CROP=1; //crop output photos?
fftw_plan fft, fftr;
double M;

#define R 500000.0		//radius curvature gaussian spherical beam
#define WAIST 1E-3		//waist of g.s. beam

#ifdef TWO_DIMENSIONAL
fftw_complex ap[N][N];
fftw_complex out[N][N];
fftw_complex old_ap[N][N];
int filter[N][N];
#endif

#ifndef TWO_DIMENSIONAL
fftw_complex ap[N][1];
fftw_complex out[N][1];
fftw_complex old_ap[N][1];
int filter[N][1];
#endif

double RS[N];

fftw_complex gamma_shift[MAX_EIGENS];
fftw_complex gamma_old, gamma_new;

struct coord
{
  int x, y;
} samples[SAMPLEPOINTS];

double rescaled_range(int start, int length);
double hurst();
void input_ap_picture (void);
void output_ap_slice (char *name);
void output_ap_picture (char *name);
void aperture_filter (void);
void propogate (double LENGTH);
void normalise_intensity_in_cavity (void);
void generate_initial_intensity ();
void scale_fft(); //correct for scale caused by FFT routines
  
void
input_ap_picture ()
{
  char buffer[200];
  int i, j, p;
  FILE *fo;

  fo = fopen ("in.raw", "r");

/*   fscanf(fo,"%s\n",&buffer);
   fscanf(fo,"%s\n",&buffer);
   fscanf(fo,"%s\n",&buffer);
   fscanf(fo,"%s\n",&buffer);
  */
  for (i = 0; i < N; i++)
    for (j = 0; j < N; j++)
      {
	fscanf (fo, "%d\n", &p);	//red
//        fscanf(fo,"%d\n",&p); //green
//        fscanf(fo,"%d\n",&p); //blue
	ap[i][j] = (double) p;
//        printf("%d %d %d\n",i,j,p);
      }

  fclose (fo);
}

void
output_ap_slice (char *name)
{
  int i,j=0;
  FILE *fo;

  #ifdef TWO_DIMENSIONAL
   j=N/2; //take slice across centre if in 2D, otherwise just read out 1D array
  #endif
  fo = fopen (name, "w");
  fprintf (fo, "#index creal cimag cabs cabs*cabs\n");
  for (i = N/2; i < N; i++)
    fprintf (fo, "%f %f %f %f %f %f %f\n", (double)(i-N/2)/((double)N*G*0.5/sqrt(2)),
	     (double)(i-N/2)/((double)N*G*0.5/sqrt(2))*sqrt(M),
	     (double)(i-N/2)/((double)N*G*0.5/sqrt(2))*M,
	     creal (ap[i][j]),
	     cimag (ap[i][j]), cabs (ap[i][j]), 
	     cabs(ap[i][j]*ap[i][j])  );
  fclose(fo);
}


void
output_ap_picture (char *name)
{
  double scr = 0.0, sci = 0.0, phi, s, v, p, q, t, f, r, g, b;
  int i, j, mag, magmax = 0, Hi;
  int size,bottom,top;
  FILE *fo;

  fo = fopen (name, "w");

//   fprintf(stderr,"Output Apperture to: %s\n",name);

   if (CROP==1) 
     {
//	size=(int)((float)N*G);
	size=N/2;
	bottom=N/4;
	top=3*N/4;
//	bottom=1+(int)((1-G)*(float)N/2);
//	top=N-(int)((1-G)*(float)N/2);
     }
   else
     {
	size=N; bottom=0; top=N;
     }
   
  fprintf (fo, "P6\n%d %d\n%d\n", size, size, 254);

  for (i = 0; i < N; i++)	//calculate scale factor
    for (j = 0; j < N; j++)
      {
	if (cabs (ap[i][j]) * cabs (ap[i][j]) > scr)
	  scr = cabs (ap[i][j]) * cabs (ap[i][j]);
/*	  if (cimag(ap[i][j])>sci)
	    sci=cimag(ap[i][j]);*/
      }

  for (i = bottom; i < top; i++)	
     //apply scale factor to normalise amount of light in cavi$
    {
      for (j = bottom; j < top; j++)	
	{
	  phi = M_PI + atan2 (cimag (ap[i][j]), creal (ap[i][j]));
//           phi=2*M_PI*(cabs(ap[i][j])*cabs(ap[i][j]))/scr;

	  s = 1.0;

	  v = (cabs (ap[i][j]) * cabs (ap[i][j])) / scr;
//           v=1.0;

//           printf("HSV: %f %f %f\n",phi,s,v);

	  //HSV->RGB formula from http://en.wikipedia.org/wiki/HSV_color_space
	  Hi = (int) (floor (phi / (M_PI / 3.0))) % 6;
	  f = phi / (M_PI / 3.0) - floor (phi / (M_PI / 3.0));
	  p = v * (1.0 - s);
	  q = v * (1.0 - f * s);
	  t = v * (1.0 - (1.0 - f) * s);

	  if (Hi == 0)
	    {
	      r = v;
	      g = t;
	      b = p;
	    }
	  if (Hi == 1)
	    {
	      r = q;
	      g = v;
	      b = p;
	    }
	  if (Hi == 2)
	    {
	      r = p;
	      g = v;
	      b = t;
	    }
	  if (Hi == 3)
	    {
	      r = p;
	      g = q;
	      b = v;
	    }
	  if (Hi == 4)
	    {
	      r = t;
	      g = p;
	      b = v;
	    }
	  if (Hi == 5)
	    {
	      r = v;
	      g = p;
	      b = q;
	    }
//           printf("%f\n",f);   

	  if (filter[i][j] != 0)	//if we're displaing the mask...
	    r = g = b = v;	//make it greyscale!
	  fprintf (fo, "%c%c%c", (int) (254.0 * r), (int) (254.0 * g),
		   (int) (254.0 * b));

//           fprintf(fo,"%d %d %d ",65535-mag,65535-mag,65535-mag);
	}
    }

  fclose (fo);

}

void
make_filter (int nsides)
{
  int i, j, n;
  float x[100], y[100];
  double grad, xinit;
  //first draw a circle
/*   for (i=0;i<N;i++) for (j=0;j<N;j++)
     if ((float)((i-N/2)*(i-N/2)+(j-N/2)*(j-N/2)) > (G*G*(float)N/2.0*(float)N/2.0))
       filter[i][j]=1; //masked
   else 
     filter[i][j]=0; //not masked*/
  //then draw polygon within circle by chopping off edges
#ifdef TWO_DIMENSIONAL
  for (n = 0; n < nsides; n++)
    {
      x[n] =
	N / 2 -
	G * (N / 2) * cos ((M_PI/(double)nsides)+ (double) n * (2.0 * M_PI / (double) nsides));
      y[n] =
	N / 2 +
	G * (N / 2) * sin ((M_PI/(double)nsides)+(double) n * (2.0 * M_PI / (double) nsides));
    }

  for (i = 0; i < N; i++)
    for (j = 0; j < N; j++)
      if (pnpoly (nsides, &x[0], &y[0], (float) i, (float) j))
	filter[i][j] = 0;
      else
	filter[i][j] = 1;
#endif
#ifndef TWO_DIMENSIONAL
   for (i = 0; i < N; i++)
     {
	
        if (i<N/2-(G*(N/2)/sqrt(2)) || i>N/2+(G*(N/2)/sqrt(2)))
          filter[i][0]=1;
        else
          filter[i][0]=0;
     }
   
#endif
}


void
aperture_filter ()
{
  int i, j=0;
  for (i = 0; i < N; i++)
#ifdef TWO_DIMENSIONAL    
     for (j = 0; j < N; j++) 
#endif
      if (filter[i][j] != 0)	//do this with unary logic
	ap[i][j] = 0.0 + 0.0I;
}

void
propogate (double LENGTH)
{
  int i, j=N/2; //set j at centre of cavity
  double shift;
   
  fftw_execute (fft); 
   
  for (i = 0; i < N; i++)	//plane wave propogation
#ifdef TWO_DIMENSIONAL
     for (j = 0; j < N; j++)
       //we need to use the modulus operator to flip the quadrants - 
       // the FFT algo changes location of the high-spectral freq
      out[i][j] *= cexp (  I * M_PI* 
	  (1.0/(double)A)*(1.0/(double)A)*
                 (double)
	  (((i + N / 2) % N - N / 2) * 
		  ((i + N / 2) % N - N / 2) +
          ((j + N / 2) % N - N / 2) * 
		  ((j + N / 2) % N - N / 2))
			  * ( LENGTH * (double)LAMBDA));
#endif

#ifndef TWO_DIMENSIONAL
      out[i][0] *= cexp (  I * M_PI* 
	  (1.0/(double)A)*(1.0/(double)A)*
                 (double)
	  (((i + N / 2) % N - N / 2) * 
		  ((i + N / 2) % N - N / 2))
			  * ( LENGTH * (double)LAMBDA));
#endif  
/*   for (i=0;i<N;i++) //Apply Hanning Window to spectral form
	  for (j=0;j<N;j++
	    out[i][j]*= 0.54-0.46 *
	               cos(2*M_PI*i/(N-1))*cos(2*M_PI*j/(N-1));
*/
  fftw_execute (fftr);
  scale_fft(); //corrects for scale in FFT algorithm
}
void lens(double f) //apply spherical lens curvature to wavefrount
  //i.e. phase retardation dependent on distance from axis
{
   int i,j=0;
   
for (i = 0; i < N; i++)
#ifdef TWO_DIMENSIONAL
  for (j = 0; j < N; j++)
     ap[i][j] *=
      cexp (I 
	    *M_PI 
	    / ((double)f*(double)LAMBDA)
	    * (((double)A/(double)N) * ((double)A/(double)N)) *
	     (double) ((i - N / 2) * (i - N / 2) +
		          (j - N / 2) * (j - N / 2)));
#endif
#ifndef TWO_DIMENSIONAL
     ap[i][0] *=
      cexp (I 
	    *M_PI 
	    / ((double)f*(double)LAMBDA)
	    * (((double)A/(double)N) * ((double)A/(double)N)) *
	     (double) ((i - N / 2) * (i - N / 2) ));   
#endif
}

void
normalise_intensity_in_cavity ()
{
  double sc = 0.0;
  int i, j=0;

  for (i = 0; i < N; i++)	//calculate scale factor
#ifdef TWO_DIMENSIONAL
     for (j = 0; j < N; j++)
#endif
       if (filter[i][j]==0) //if within aperture
      sc += cabs (ap[i][j])*cabs(ap[i][j]);

   sc=(double)sc; //discard imaginery part
#ifdef TWO_DIMENSIONAL
   sc=sc/((double)N*(double)N*G*G); //average abs. value of pixel in cavity
#endif
#ifndef TWO_DIMENSIONAL
   sc=sc/((double)N*G);
#endif
   sc=sqrt(sc); //take sqrt to get 
//   fprintf(stderr,"Normalise intensity: sc %f\n",sc);
   
  for (i = 0; i < N; i++)	//apply scale factor to normalise amount of light in cavity
#ifdef TWO_DIMENSIONAL
     for (j = 0; j < N; j++)
#endif
       ap[i][j]=ap[i][j]/sc; 
}

void
generate_initial_intensity ()
{
  int i, j=0;

  for (i = 0; i < N; i++)
#ifdef TWO_DIMENSIONAL
     for (j = 0; j < N; j++)
#endif
	{
//           ap[i][j]=(i-N/2)+((j-N/2)*I);

/*	     if (i>0.25*N
		 && i<0.75*N
		 && j>0.25*N
		 && j<0.75*N )
	          ap[i][j]=1.0+0.0I;
	     else
	       ap[i][j]=0.0+0.0I;
  */
	  ap[i][j] = cexp (
//                  -I*2*M_PI/LAMBDA*
//                  ((double)((i-N/2)*(i-N/2)+(j-N/2)*(j-N/2))/(double)(N))
//                  /R 
			    - 
			   (A*A)* ((double)
			     ((i - N / 2) * (i - N / 2) +
			      (j - N / 2) * (j - N / 2)) 
				   / (double) (N * N)) /
			    (WAIST*WAIST));
	  ap[i][j] = 1.0 + 0.0I;	//DIRTY! :)

//           fprintf(stderr,"i: %d j: %d\t%f + %f i\n",i,j,creal(ap[i][j]),cimag(ap[i][j]));
	}

}


void scale_fft() //correct for scale caused by FFT routines
{
   int i,j;
     for (i = 0; i < N; i++)
#ifdef TWO_DIMENSIONAL
       for (j = 0; j < N; j++)
           ap[i][j]/=(double)N*(double)N; //correct for scaling of FFT algo
#endif
#ifndef TWO_DIMENSIONAL
       ap[i][0]/=(double)N; //correct for scaling of 1D FFT algo
#endif
}

void
output_filter ()
{
  int i, j;
  for (i = 0; i < N; i++)
    for (j = 0; j < N; j++)
      ap[i][j] = 1.0;
  aperture_filter ();
}

fftw_complex calculate_gamma()
//Calculate Gamma shift factor by comparing average of succesive pixels once stabilised on eigenmode
{  
	int i,j=0,ap_points = 0;
   fftw_complex gamma_new=0.0;
	    for (i = 0; i < N; i++)
#ifdef TWO_DIMENSIONAL
     for (j = 0; j < N; j++)
#endif
		if (filter[i][j] ==0 && cabs(ap[i][j])>0.05)
		  {
		    ap_points++;
		    gamma_new += ap[i][j] / old_ap[i][j];
		  }
	    gamma_new /= (double) ap_points;
//   fprintf(stderr,"%d",ap_points);
   return(gamma_new);
}

void apply_gamma_shift(int shift)
{
   int i,j=0;
		 for (i = 0; i < N; i++)
#ifdef TWO_DIMENSIONAL
     for (j = 0; j < N; j++)
#endif
		    ap[i][j] -= old_ap[i][j] * gamma_shift[shift];	//rotate between gamma_shift shifts on each successive pass     
		//So - just what are we meant to do here? Apply subtractive shift to previous frame?
//         fprintf(stderr,"Apply Gam: %f + %f I\n",creal(gamma_shift[1+passes%eigenmode_count])
//                 ,cimag(gamma_shift[1+passes%eigenmode_count]));
}


double hurst(char *name)
  //Derived from equations at
  // http://www.bearcave.com/misl/misl_tech/wavelets/hurst/index.html
{
  FILE *fo;  
   int i=0,n;
   int bits=0,size=512;
 
   double data[20][2];
   
   double avg;
   double extent=0.4; //fraction of guard band we'll be using
   size=(int)(extent*G*N)/2.0;
   printf("%d\n",size);
 
   fo = fopen (name, "w");   
 
   bits=size;
   
   while (bits>=8)
     {
	avg=0.0;
	for (n=0;n<size;n+=bits)
	     avg+=rescaled_range(n+N/2,bits);
	
	avg/=(double)(size/bits); //was size/bits
	
	fprintf(fo,"%f %f %f %f\n",1.0/(double)bits,avg,log((double)bits)/log(10.0),log(avg)/log(10.0));
	
	data[i][0]=log((double)bits)/log(10.0); data[i][1]=log(avg)/log(10.0);
	
	i++;
	bits/=2;
     }
	
	fprintf(fo,"#D: Approx: %f\n",(data[0][1]-data[i-1][1])/(data[0][0]-data[i-1][0]) );
   fclose(fo);
   
   return(0.0); //FIXME
}

double rescaled_range(int start, int length)
  //Dervied from equations at
  // http://www.bearcave.com/misl/misl_tech/wavelets/hurst/index.html
{
   int i;
   double avg=0.0,min=0.0,max=0.0,sd=0.0;
   
   for (i=start;i<start+length;i++)
     avg+=(double)cabs(ap[i][0]);
   avg/=(double)length; //calculate avg. value
   
   RS[0]=0.0;
   for (i=1;i<=length;i++)
     {	
     RS[i]=RS[i-1]+(double)cabs(ap[i+start][0])-avg;
     if (RS[i]<min) min=RS[i];
     if (RS[i]>max) max=RS[i];
     }
   for (i=start;i<start+length;i++)
     sd+=((double)cabs(ap[i][0])-avg)*((double)cabs(ap[i][0])-avg);

   sd/=(double)length; //now contains variance
   sd=sqrt(sd); //now S.D.
   
  return ((max-min)/sd);
}



main ()
{
  int i, j=0, k, l, passes, n, eigenmode_flag, eigenmode_count, ap_points,shift;
   int framecount=0;
  char name[100],tmp[100];
  double tmpr, tmpi, sc, total_error,Neq,conjugate_plane;
   double gimag,greal;
   double g1,g2,FOCAL_CONVERSION; //g-factors for laser cavity


#ifdef TWO_DIMENSIONAL
  fft = fftw_plan_dft_2d (N, N,
  			  &ap[0][0], &out[0][0], FFTW_FORWARD, FFTW_ESTIMATE);
  fftr = fftw_plan_dft_2d (N, N,
			   &out[0][0],
			   &ap[0][0], FFTW_BACKWARD, FFTW_ESTIMATE);
#endif
   
#ifndef TWO_DIMENSIONAL
  fft = fftw_plan_dft_2d (N, 1,
  			  &ap[0][0], &out[0][0], FFTW_FORWARD, FFTW_ESTIMATE);
  fftr = fftw_plan_dft_2d (N, 1,
			   &out[0][0],
			   &ap[0][0], FFTW_BACKWARD, FFTW_ESTIMATE);
#endif   
  srand (time (NULL));

  fprintf (stderr, "Plans created...N:%d\n",N);

// for (M=1.5;M<1.6;M+=0.6)   
  for (n = 5; n < 6; n++) //n-sided polygon for aperture
//    for (FOCAL = -2.0; FOCAL > -20.0; FOCAL -= 1.0)
      {
//	 g1= -1.0526; //-1.01; //-1.055;
	 g1=-1.01;
//	 g1=-1.002;
//	 M=1.9;
//	 g1=(M+1.0) /(2.0*M);
	 g2=g1;
	 printf("g1=%f g2=%f\n",g1,g2);
//	 FOCAL=(-M*L)/((M-1)*(M-1));
//	 FOCAL=0.225;	
	 FOCAL_CONVERSION=-g2*L/(g2-1);
	 
	 FOCAL=1/(2-2*g1);

	 conjugate_plane=(L/2)*sqrt((g1+1)/(g1-1)); //distance x from centre of cavity
	 M=(-g1+sqrt(g1*g1 - 1))/(-g1-sqrt(g1*g1-1)); //Magnification of cavity
	 Neq=12.0;
	 LAMBDA=((M*M)-1)/(2*M)*(A*G*A*G/8.0)/(L*Neq); //choose lambda from previous Neq
	 fprintf(stderr,"Lamda: %e Neq: %f\n",LAMBDA,Neq);
	 Neq=((M*M)-1)/(2*M) * (A*G*A*G/8.0)/(L*LAMBDA); //calculate Neq from A,Lambda,L & M
	fprintf(stderr,"Congjugate planes: u:%f v:%f M: %f x: %f\n",L/2-conjugate_plane,L/2+conjugate_plane,M,conjugate_plane);

	 //EQUIVALENT LENSGUIDE CONVERSIONS
//	 FOCAL=-(g2*L)/(2*(g1*g2-1)); //focal length of equiv lensguide - Eqn 16, GHCN notes	 
//	 L=2*g1*L; //equivalennt freescale length - Eqn 15, GHCN notes
	 


	 fprintf(stderr,"M: %f L: %f Focal: %f Focal_Conversion %f N: %f Neq: %f\n",
		 M,L,FOCAL,FOCAL_CONVERSION,
		 (0.5*A*G*0.5*A*G/2.0)/(L*LAMBDA),
		 //((1-L/FOCAL)-1)/2.0   *(0.5*A*G*0.5*A*G/2.0)/(L*LAMBDA));
		 Neq);
//	 ((M-1)/2.0 * (A*G*A*G/8.0)/(L*LAMBDA)));
 
		 
//   for (i=0;i<N;i++) for (j=0;j<N;j++)
	//    shift[i][j]=0.0+0.0I;    
	make_filter (n);
	fprintf (stderr, "Npolygon: %d M: %f Focal: %f\n", n, M,FOCAL);

//   for (L=0.001;L<=0.024;L+=0.001) //10 240 10
//     {
//      fprintf(stderr,"Going for Length %f\n",L);


//   input_ap_picture(); //Lena
//   
	gamma_old = gamma_new = 0.0 + 0.0I;

	for (i = 0; i < N; i++)
#ifdef TWO_DIMENSIONAL
	   for (j = 0; j < N; j++)
#endif
	    old_ap[i][j] = 0.0 + 0.0I;
	 
//	 for (greal=-1.0;greal<1.0;greal+=0.1)
//	   for (gimag=-1.0;gimag<1.0;gimag+=0.1)
//	     {
	eigenmode_count = 0;		
//	     gamma_shift[0]=greal+gimag*I;
		
	generate_initial_intensity ();
	 
//	        sprintf(name,"%.10d.pnm",0);
//                output_ap_picture(name);
//	     sprintf(name,"%.10d.log",0);
//		output_ap_slice(name);	     
	 
	aperture_filter ();

//	        lens(-FOCAL_CONVERSION);		 
	 for (passes = 0; passes < 10000; passes++)
	  {
//      fprintf(stderr,"Nsides: %d Passes %d\n",n,passes);
//        sprintf(name,"%.10d.pnm",framecount++);
//        output_ap_picture(name);	     

	     //EQUIV LENSGUIDE
//	        lens(FOCAL);	     
//		propogate (L);
		
	     lens(FOCAL);
             propogate(L/2-conjugate_plane);
	     aperture_filter();
             propogate(L/2+conjugate_plane);	     
	     lens(FOCAL);
	     
	     propogate(L/2+conjugate_plane);
	     aperture_filter();
	     propogate(L/2-conjugate_plane);

	     //propogate(L);

  	      //Gamma shift application
	      //Start of SHIFT selection
	     /* the following code applies the shifts in a straight series 
	      * with a gap of SPACERS between each rotated application.
	      * So it looks like abc......abc.....abc.... etc.
	      */
	     
	      shift=(passes+(SPACERS-1))%(SPACERS+eigenmode_count) - SPACERS;
	      
//	      if (shift>=0)
//	       shift=eigenmode_count-shift-1;

//	     fprintf(stderr," %d ",shift);
	     /* the following code applies the shifts in rotation, spaced by
	      * a gap of SPACERS between each single application of a shift.
	      * So it looks like a......b......c......a......b......c.... etc.
	      */
/*	     if (passes%SPACERS==0 && eigenmode_count>0)
	       shift=(passes/SPACERS)%eigenmode_count;
	     else
	       shift=-1;
*/
	     //End of SHIFT selection
	      if (shift<0) { fprintf(stderr,"."); sprintf(tmp,"X");}
	      else 
		{
		   fprintf(stderr,"%c",'a'+shift);
		   
		   for (i=1;i<=shift+1;i++)
		     tmp[i]='A'+i-1;
		   tmp[shift+2]=0;
		   
		   apply_gamma_shift(shift);
		}
		              
	    gamma_new = calculate_gamma();
	     framecount++;
//	        sprintf(name,"%.10d_%s_Mode:%d_G:%f+%fI.pnm",framecount,tmp,eigenmode_count,creal(gamma_new),cimag(gamma_new));
//                output_ap_picture(name);
// 	    normalise_intensity_in_cavity ();	                     
//	     sprintf(name,"%.5d.pnm",framecount);
  //              output_ap_picture(name);	     
//		output_ap_slice(name);	     
//	     exit(-1);


/*      printf("G_new %f + %fI cabs: %f old:new  %f\n",
              creal(gamma_new),cimag(gamma_new),
	     cabs(gamma_new),
              cabs(gamma_new-gamma_old)
	     );
*/

          if ( passes>15 && cabs (gamma_new - gamma_old) < (double) TOLERANCE)     //see if stabailised to eigenmode by non-varying Gamma shift
	      {
		fprintf (stderr, "c@%d\n", passes);
//        fprintf(stderr,"Convergence to Eigenmode, with %f Tolerance.\n",TOLERANCE);

		gamma_shift[eigenmode_count] = gamma_new;	//save gamma into shift table

		printf ("Avg Gamma[%d]: %f + %fI\tAbs:%f\n", eigenmode_count,
			creal (gamma_shift[eigenmode_count]),
			cimag (gamma_shift[eigenmode_count]),
			cabs(gamma_shift[eigenmode_count]));
		 
	     sprintf(name,"%dr.log",eigenmode_count);		 
		 output_ap_slice(name);

		 sprintf(name,"h%d.log",eigenmode_count);
		 hurst(name);
		 
		 //remove the following cludge
/*	        lens(FOCAL);
		propogate (L);		 
	        lens(FOCAL_CONVERSION);		 	        
		propogate (-(0.5-conjugate_plane));
*/
//		 lens(FOCAL_CONVERSION); //back to full cavity
		 lens(1/(2-2*g1)); //FOCUS as actually mirror
		 propogate(0.5+conjugate_plane); //propogate forwards

		 normalise_intensity_in_cavity ();	     
		 
	         sprintf(name,"%du_lr.log",eigenmode_count);		 
		 output_ap_slice(name);
		 
		 sprintf(name,"hc%d.log",eigenmode_count);
		 hurst(name);


#ifdef TWO_DIMENSIONAL //if making a 2D eigenmode, output the pretty eigenmode!
		sprintf(name,"%du_lr.pnm",eigenmode_count); 
		output_ap_picture (name);
#endif

		 aperture_filter();
		 propogate (2*conjugate_plane);		 
		 sprintf(name,"%dv_lr.log",eigenmode_count);		 		 
		 output_ap_slice(name);
		 
		 propogate(0.5-conjugate_plane);
		 lens(1/(2-2*g1)); //FOCUS as actually mirror
		 propogate(0.5-conjugate_plane);
		 sprintf(name,"%dv_rl.log",eigenmode_count);		 		 
		 output_ap_slice(name);

		 propogate (2*conjugate_plane);		 
		 sprintf(name,"%du_rl.log",eigenmode_count);		 		 
		 output_ap_slice(name);		 
#ifdef TWO_DIMENSIONAL //if making a 2D eigenmode, output the pretty eigenmode!
		                 sprintf(name,"%du_rl.pnm",eigenmode_count);
		                 output_ap_picture (name);
#endif
		 
		eigenmode_count++;	//keep count of already discovered eigenmodes
		passes = 0;	//reset passes so we have full range to settle to next mode
		 gamma_old=gamma_new=0.0+0.0I; //reset gamma factors
	      }
	     
  	    aperture_filter ();     
 	    normalise_intensity_in_cavity ();	     
	     
	    for (i = 0; i < N; i++)
#ifdef TWO_DIMENSIONAL
	       for (j = 0; j < N; j++)
#endif
		old_ap[i][j] = ap[i][j];
//      memcpy(old_ap,ap,sizeof(fftw_complex)*N*N);
	     
	    gamma_old = gamma_new;
	    
	    if (eigenmode_count >= MAX_EIGENS)	//once we've gathered this many modes
	      break;		//break out the for-loop!

	  }
	fprintf (stderr, "Reset\n");
//	     }
	 
//      sprintf(name,"npoly%d_Foc%f_passes%.5d.pnm",n,FOCAL,passes);
//      output_ap_picture(name);


//        output_ap_slice((int)(L*1000));       
      }
  fftw_destroy_plan (fft);
  fftw_destroy_plan (fftr);
}
\end{lstlisting}

\clearpage
\section{Video Fractal Codes}
\lstset{basicstyle=\scriptsize,breaklines=true}
\begin{lstlisting}
/* Jarvist Frost 2004-2006
 * Program to create 'video-fractals'
 */

//#include <file.h>
#include <stdio.h>
#include <math.h>
#include <limits.h>

#define MAG (1.4)
#define X_RES 500
#define Y_RES 500

#define X_OFF 0 //offset of newcenter in pixels
#define Y_OFF 0

#define TWIST 3.14/6 //radians twist between zoom's

#define PIXW 0.65 //0.65 //width of sensor pixel in display pixels 

#define BACKGROUND 140

int curpic[X_RES][Y_RES];
int newpic[X_RES][Y_RES];
void outputpic(char *filename);
void inputpic(char *filename);

main()
{
   int x,y,loops;
   char filename[20];

   //fill display with white noise
   srand(123);
   for (x=0;x<X_RES;x++)
     for (y=0;y<Y_RES;y++)
       curpic[x][y]=rand();
   
   inputpic("begin.pgm");


   
   outputpic("first.pgm");
//   zoom();
//   swap();
//   outputpic("test2.pgm");
   
   for(loops=0;loops<150;loops++)
     {
	printf("%d\n",loops);
	sprintf(filename,"pic%.3d.pgm",loops);
	if (loops%10==0)
  	   outputpic(filename);

	zoom(); swap();
     }
   
   outputpic("last.pgm");
}

swap()
{
  int x,y,max=0;
   double light;
   
   for (x=0;x<X_RES;x++)
     for(y=0;y<Y_RES;y++)
       {  
       curpic[x][y]=newpic[x][y];
       if (curpic[x][y]>max)
          max=curpic[x][y];
       }

   for (x=0;x<X_RES;x++)
     for(y=0;y<Y_RES;y++)
       curpic[x][y]*=INT_MAX/max;   
}


zoom()
{
 int x,y;
   double nx,ny,np,dx,dy,r,theta,phi;
   
   for (x=0;x<X_RES;x++)
     for(y=0;y<Y_RES;y++)
       {
	  dx=(0.5+(double)(x-(X_RES/2)))/MAG;
	  dy=(0.5+(double)(y-(Y_RES/2)))/MAG;
	  
	  r=sqrt(dx*dx+dy*dy);
	  theta=atan2(dy,dx);
	  phi=theta+TWIST;
	  
	  ny=r*sin(phi);
	  nx=r*cos(phi);
	  
	  nx=nx+(double)(X_RES/2+X_OFF);
	  ny=ny+(double)(Y_RES/2+Y_OFF);
	  
//	  printf("x: %d nx: %f y: %d ny: %f\n",x,nx,y,ny);
	  
	  np=0; 
	  np+=vo((int)nx+1,(int)ny+1) *(nx-(double)((int)nx))*(ny-(double)((int)ny)); //top-right pixel
	  np+=vo((int)nx,(int)ny+1) *(PIXW-(nx-(double)((int)nx)))*(ny-(double)((int)ny)); //top-left pixel
	  np+=vo((int)nx+1,(int)ny) *(nx-(double)((int)nx))*(PIXW-(ny-(double)((int)ny))); //bot-right pixel
	  np+=vo((int)nx,(int)ny) *(PIXW-(double)(nx-(double)((int)nx)))*(PIXW-(ny-(double)((int)ny))); //bot-left pixel
	  
	  np*=1.0/(PIXW*PIXW); //compensates for size of pixel otherwise 'losing' light from the feedback
	  
	  newpic[x][y]=(int)np;
//	  printf("np: %f\n",np);
       }
   
   
}

int vo(int x, int y) //value of a particular pixel; with bounds checking
{
   if (x<0||x>X_RES) return(BACKGROUND);
   if (y<0||y>Y_RES) return(BACKGROUND);
   
   return (curpic[x][y]);	
}
	

void outputpic(char *filename)
{
   int x,y;
   FILE * f;
   f=fopen(filename,"w");
   
   fprintf(f,"P5 %d %d 255\n",X_RES,Y_RES); //.pgm filetype - binary form
   
   for (x=0;x<X_RES;x++)
      for (y=0;y<Y_RES;y++)
         fprintf(f,"%c",curpic[x][y]/(INT_MAX/255));
   
   fclose(f);
}

void inputpic(char *filename)
{ 
  int x,y;
   int tmp;
   
   FILE * f;
   f=fopen(filename,"r");

   fscanf(f,"P2 500 500\n"); //.pgm filetype

   for (x=0;x<X_RES;x++)
        for (y=0;y<Y_RES;y++)
       {
             fscanf(f,"%d",&tmp);
//	  printf("%d\n",curpic[x][y]);
// printf("tmp: %d\n",tmp);
	  curpic[x][y]=tmp*(INT_MAX/255);
       }
   

   fclose(f);
}
\end{lstlisting}

\end{document}